%% file: traj_comp_long.tex
\documentclass{sig-alternate-05-2015}

\usepackage{subfig}
\usepackage{algorithm}
\usepackage{algorithmic}

\usepackage{graphicx}
\usepackage{balance}  
\usepackage{subfig}
\usepackage{algorithm}
\usepackage{algorithmic}
\usepackage{url}
\newtheorem{thm}{Theorem}

\newtheorem{mydef}{Definition}
\DeclareMathOperator*{\argmax}{arg\,max}

\begin{document}
%

\title{Prediction-based Online Trajectory Compression}
%
%
%
%
%

\numberofauthors{1} 
%
\author{
\alignauthor
{
Arlei Silva{\small $~^{\dagger}$}, Ramya Raghavendra{\small $~^{*}$}, Mudhakar Srivatsa{\small $~^{*}$}, Ambuj K. Singh{\small $~^{\dagger}$} }\vspace{1.6mm}\\
$^{\dagger}$\affaddr{Computer Science Department, University of California, Santa Barbara, CA, USA}\\
\vspace{1.6mm}
$^{*}$\affaddr{IBM T.J. Watson Research Center, Yorktown Heights, NY, USA}\\
\email{\{arlei,ambuj\}@cs.ucsb.edu,\{rraghav,msrivats\}@us.ibm.com}\\
}

\maketitle

\maketitle
\input{0_abstract}

\input{1_introduction}

\input{3_trajectory_data_management}

\input{4_real_time_compression}

\input{5_experimental_results}

\input{2_related_work}

\input{6_conclusion}

\bibliographystyle{abbrv}
\bibliography{traj_comp_long}  

\appendix
\input{appendix}

\end{document}

%% file: 0_abstract.tex
\begin{abstract}
Recent spatio-temporal data applications, such as car-shar\-ing and smart cities, impose new challenges regarding the scalability and timeliness of data processing systems. Trajectory compression is a promising approach for scaling up spatio-temporal databases. However, existing techniques fail to address the online setting, in which a compressed version of a trajectory stream has to be maintained over time. In this paper, we introduce ONTRAC, a new framework for map-matched online trajectory compression. ONTRAC learns prediction models for suppressing updates to a trajectory database using training data. Two prediction schemes are proposed, one for road segments via a Markov model and another for travel-times by combining Quadratic Programming and Expectation Maximization. Experiments show that ONTRAC outperforms the state-of-the-art offline technique even when long update delays (4 minutes) are allowed and achieves up to 21 times higher compression ratio for travel-times. Moreover, our approach increases database scalability by up to one order of magnitude.\\
\textbf{Categories and Subject Descriptors:} H.2.8 [\textbf{Database Management}]: Database applications $-$ \textit{data mining} \\
\textbf{General Terms:} Algorithms, Experimentation \\
\textbf{Keywords:} Trajectory, Compression, Online
\end{abstract}

%% file: 1_introduction.tex
\section{Introduction}
\label{section::introduction}

\begin{figure}[ht!]
\centering
\subfloat[Full]
{
\includegraphics[keepaspectratio, width=0.2\textwidth]{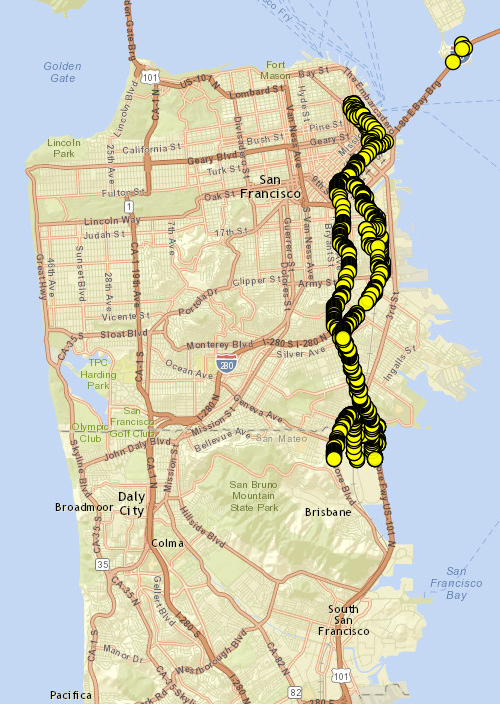}
}
\subfloat[Compressed]
{
\includegraphics[keepaspectratio, width=0.2\textwidth]{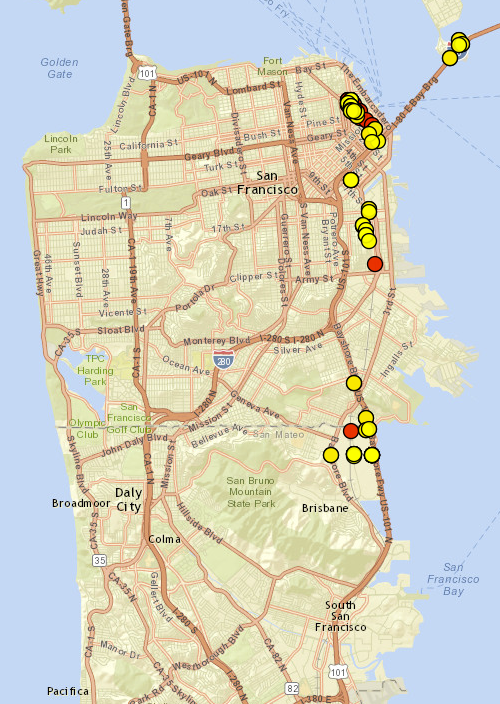}
}
\caption{Full trajectory vs. ONTRAC compression for a real cab trajectory. The compression ratio achieved is $19.6$.\label{fig::ramya_figure}}
\end{figure}

Online Trajectory compression is a promising approach for enabling the processing of trajectory data at a scale that meets the demands of modern applications. However, existing work \cite{Cao:2005:NMI:2131560.2131576,kellaris2013map,pvldbSongSZZ14} fails to address the online setting, in which an accurate compressed version of a trajectory stream has to be maintained as new updates arrive. For instance, car-sharing apps, such as \textit{Uber} and \textit{Lyft}, have to handle online trajectory data at a massive scale in order to perform user-driver assignment, improve routing algorithms and protect both drivers' and passengers' safety. Moreover, \textit{megacities} like Beijing are investing in traffic analytics as means to reduce congestions and air pollution. An important aspect shared by all these applications is that trajectories can be map-matched, since vehicles move mostly over a road structure. Existing work has shown that we can make use of such knowledge to improve trajectory data processing \cite{Cao:2005:NMI:2131560.2131576}.

\begin{figure*}[ht!]
\centering
\includegraphics[keepaspectratio, width=0.8\textwidth]{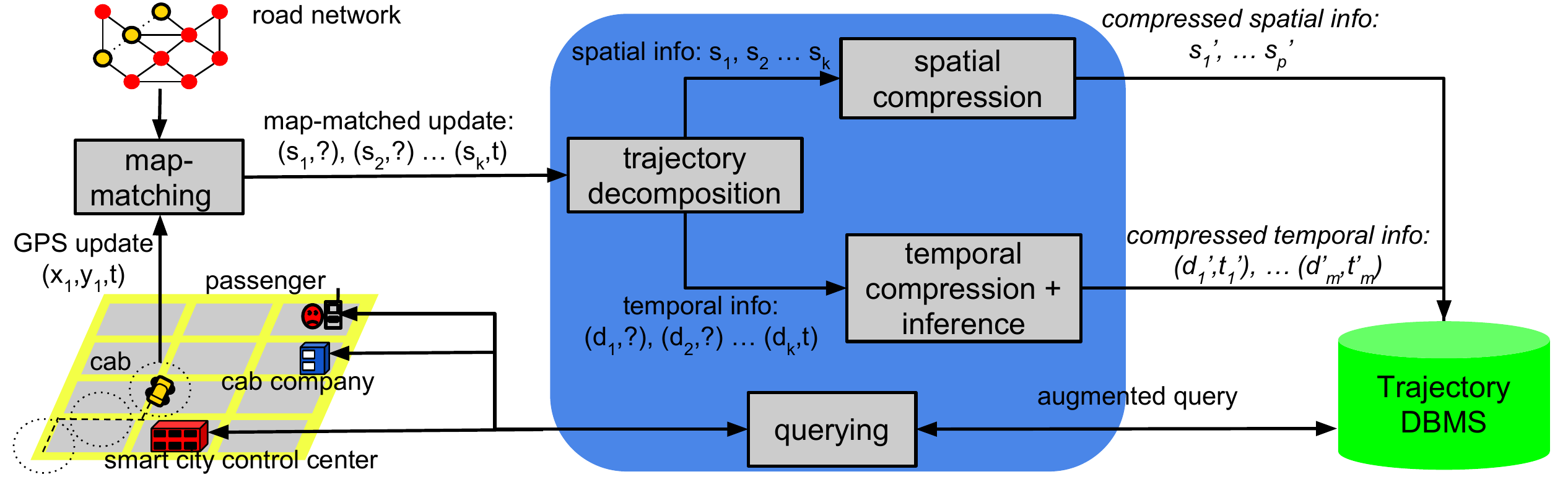}
\caption{Application scenario: Online trajectory compression for smart cities. \label{fig::ontrac}}
\end{figure*}

Figure \ref{fig::ontrac} illustrates an online trajectory compression application scenario motivated by \textit{smart cities}. Cab trajectory data is collected using GPS devices and stored in a trajectory DBMS in the \textit{cloud}. A cab updates its trajectory in the database by sending its (lat-long) coordinates every minute. Recent GPS updates are map-matched to the road network structure and then processed by an online trajectory compression framework, which inserts a compressed version of the trajectory into the database. Several applications query the DBMS for different purposes, including a cab-dispatching app, the cab company, and a smart city control center. Each query is augmented by the trajectory compression framework as means to become compatible with the compressed data schema. The retrieved results are then uncompressed by the framework before they are sent to the applications. For instance, the smart city control center might compare recent cab speeds against historical data to detect traffic jams. The focus of this paper is on how to scale these applications for large cities, with tens of thousands of cabs.

The natural solution to the online trajectory compression problem is to extend state-of-the art offline schemes \cite{Cao:2005:NMI:2131560.2131576,kellaris2013map,pvldbSongSZZ14} using a caching mechanism with bounded delays. If delays are short (e.g. in the order of one minute), the trajectory database might still answer queries in an ``almost online'' fashion. However, our experiments show that the compression ratio achieved by an offline approach degrades drastically as the size of the accepted delay window decreases. Also, regularly \textit{flushing} this cache to the spatial index, which typically can handle a few thousand inserts/second (see Section \ref{subsec::exp_querying_storage}), might deteriorate the performance of the system.

In this paper, we introduce a new framework for online trajectory compression called \textbf{ONTRAC} (ONline TRAjectory Compression).
ONTRAC's design started from the following open question: \textit{Can we apply trajectory prediction as means to suppress online updates to a large trajectory database?} While there is an extensive body of research on trajectory prediction in the literature \cite{xue2013destination,hunter2009path,ide2011trajectory,kim2007path,yang2013travel,wang2014travel,yuan2011driving}, none of these addresses the online compression task. 

Our approach decomposes trajectories into two components ---spatial and temporal trajectory information--- for which different prediction algorithms are proposed. Whenever a given update is correctly predicted, it is suppressed from the database. By combining compressed trajectory information and the prediction model, ONTRAC can accurately recover these suppressed updates at query time. We show one example of a real cab trajectory compressed using ONTRAC in Figure \ref{fig::ramya_figure}. The compression ratio achieved for this trajectory is $19.6$ (i.e. $95$\% of the updates are suppressed). The motivation for ONTRAC comes from information theory, where, dictionary-based compression (e.g. Huffman coding) requires knowledge of the complete string to be compressed but prediction-based schemes (e.g. prediction by partial matching \cite{begleiter2004prediction,feder1994relations}) compress new symbols using a prediction model. However, in typical trajectory compression applications, compressed data is stored into a database with query processing capabilities. Information theory is mostly focused on bit string representations that cannot be processed by existing database technology.

ONTRAC compresses spatial trajectory information, i.e. the sequence of road segments visited by an object, by predicting segment updates using a Markov model. For temporal compression, ONTRAC combines \textit{Quadratic Programming} (QP) \cite{boyd2004convex} and \textit{Expectation-maximization} (EM) \cite{dempster1977maximum} to learn a prediction model for travel-times. Our approach handles the sparseness and noise of GPS updates, a problem we call \textit{temporal trajectory inference}, by computing maximum-likelihood estimates for missing temporal trajectory information. Such estimates are based on a Gaussian model for travel-times that is fitted using a convex QP formulation. We illustrate how data compressed using ONTRAC can be efficiently queried --with partial trajectory decompression-- for a typical historical query that asks for the location of a given vehicle at a particular instant of time.

We summarize our main contributions as follows:
\begin{itemize}
\item We introduce the online map-matched trajectory compression problem in databases, which consists of minimizing the number of updates in a trajectory stream.
\item We present ONTRAC, an online trajectory compression framework that supports lossless spatial compression and error-bounded temporal compression.
\item We demonstrate the effectiveness of ONTRAC using real datasets. While our approach is fully online, the baseline requires long update delays (4 minutes) to achieve the same spatial compression ratio. Moreover, ONTRAC obtains up to 21 times higher compression ratio for temporal trajectory information. We also show how ONTRAC's compression increases database scalability by up to one order of magnitude. 
\end{itemize}

%% file: 3_trajectory_data_management.tex
\section{Online Trajectory Compression}
\label{sec::online_trajectory_compression_and_inference}

A road network is a directed graph $G(V,E)$ where $V$ is a set of road segments and $(s_i,s_j) \in E$ iff an object can navigate from $s_i$ to $s_j$ without passing through any other segment in $V$ \cite{ide2011trajectory}. A GPS trajectory is a sequence of triples $<(x_1,y_1,t_1) \ldots (x_q,y_q,t_q)>$, where $x_i$ is a longitude, $y_i$ is a latitude and $t_i$ is a timestamp. GPS trajectories can describe any movement on Earth's surface. However, this paper assumes that trajectories are constrained to the road network $G$, thus we can represent a GPS trajectory as a map-matched trajectory $<(s_1,t_1) \ldots (s_r,t_r)>$, where $s_i \in V$ and $t_i$ is the time when $s_i$ was traversed. 

Since GPS updates are sparse, map-matched trajectories have missing temporal information. An update $(s_i,?)$ for which the timestamp $t_i$ is unknown is called a missing update. The temporal trajectory inference problem consists of inferring such information based on the remaining updates. 

\begin{mydef}
\textbf{Temporal Trajectory Inference Problem:} Given a map-matched trajectory $T'=<(s_1,t_2), (s_2,?)$ $\ldots (s_m,t_m)>$ with missing temporal information, compute estimates for the timestamps $t_i$ of the missing updates $(s_i,?)$.
\end{mydef}

\begin{figure}[ht!]
\centering
\subfloat[Road Network\label{fig::road_net}]{
\includegraphics[keepaspectratio, width=0.45\textwidth]{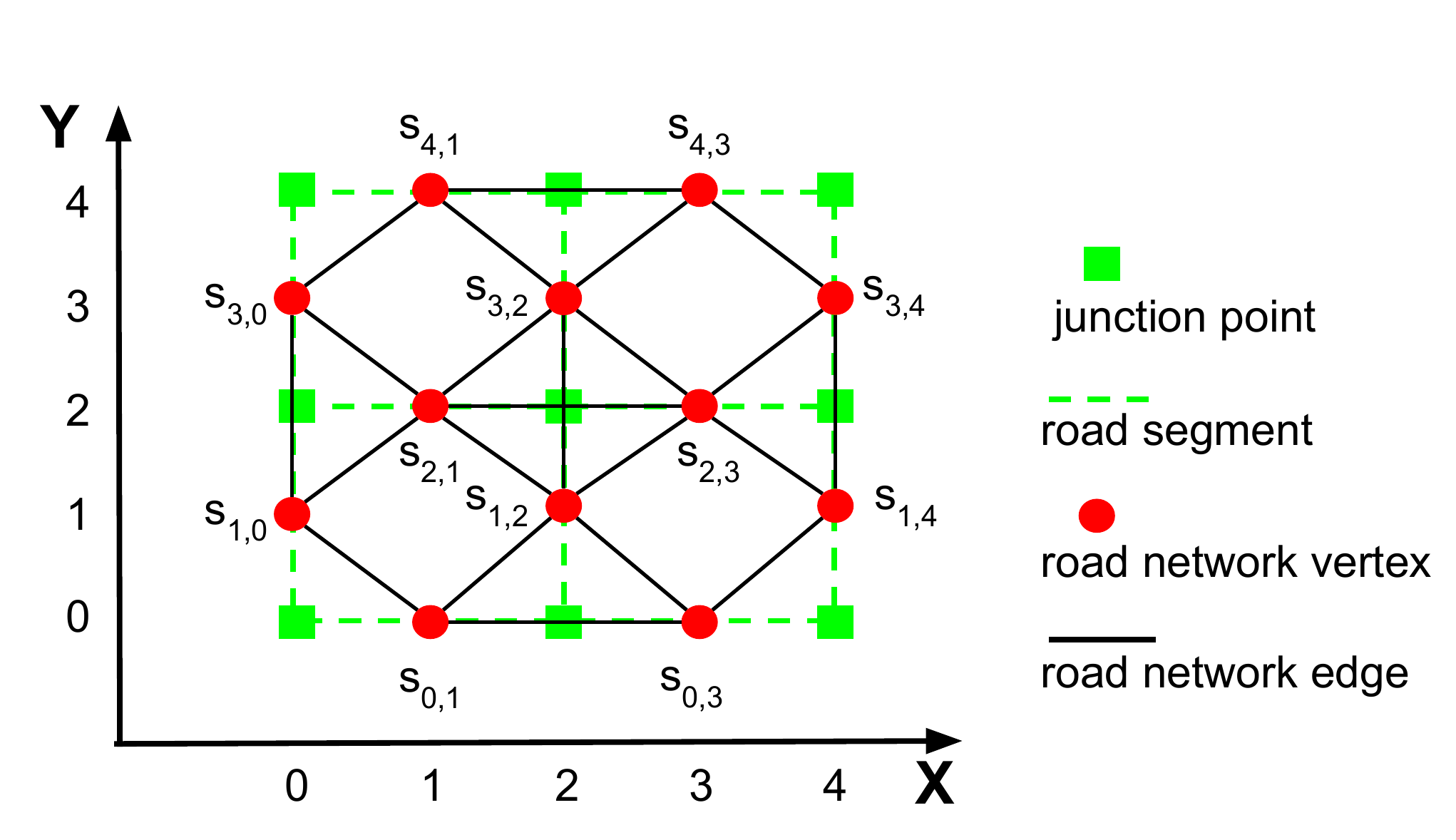}
}

\subfloat[Trajectory stream \label{table:example_trajectories}]{
\begin{tabular}{| c  c |}
\hline
\textbf{id}& \textbf{trajectory}\\
$o_1$ & $<(s_{2,1},10),(s_{2,3},20),(s_{3,4},40)>$\\
$o_2$ & $<(s_{1,4},5),(s_{2,3},15),(s_{3,2},30)>$\\
$o_3$ & $<(s_{3,2},?),(s_{2,3},?),(s_{1,4},17)>$\\
$o_4$ & $<(s_{1,2},10),(s_{2,1},15),(s_{2,3},?),(s_{3,4},35)>$\\
\hline
\end{tabular}
}
\caption{Running example: (a) Road network for a 2x2 grid where each road segment is 2 units long.  (b) Trajectory stream with four trajectories. \label{fig::running_example}}
\end{figure}

A map-matched trajectory stream is a sequence of updates $(o_i,s_i,t_i)$ ordered by $t_i$, where $o_i$ is an object, $s_i$ is a road segment and $t_i$ is a timestamp. Nevertheless, for sake of simplicity, our algorithms will be described in terms of aggregated trajectories. Each object $o_i$ has an associated trajectory $T_i = <(s_1,t_1) \ldots (s_m,t_m)>$ that aggregates all its updates $(o_i,s_1,t_1) \ldots (o_i,s_m,t_m)$. Notice that these trajectories might have missing temporal information. A trajectory stream $\mathcal{D}$ is a set of trajectories $\{T_1, \ldots T_n\}$ that are processed in an online manner (i.e. one update at a time). 

Given a trajectory stream $\mathcal{D}$, the online trajectory compression problem consists of keeping a compressed approximated representation $\mathcal{D}'$ that enables the recovery of the trajectories in $\mathcal{D}$ over time with some guarantees. In particular, we consider compression schemes that are \textit{spatially lossless} and \textit{temporally lossy}. A spatially lossless trajectory compression scheme, guarantees exact recovery of the sequence of road segments $<s_1, s_2 \ldots s_k>$ of any trajectory in $\mathcal{D}$. Moreover, a temporally lossy compression, guarantees that for any sequence of segments, the timestamp $t_i$ associated with a segment $s_i$ will be recovered as an estimate $t_i'$ with bounded error ($|t_i-t_i'| < \lambda$).

\begin{mydef}
\textbf{Online Trajectory Compression Problem:} Given a map-matched trajectory stream $\mathcal{D}$, maintain a compressed stream $\mathcal{D}'$ that is spatially lossless and temporally lossy with error bounded by $\lambda$ over time.
\end{mydef}

Figure \ref{fig::road_net} and Table \ref{table:example_trajectories} show a road network and a trajectory stream, respectively, used as a running example throughout this paper. The road network is built based on a 3x3 grid over a 2-dimensional surface with latitude (Y) and longitude (X) spanning the range [0,4]. Road segments are 2 units long and their IDs are set according to their respective location in the grid. The trajectory stream covers 4 objects ($o_1$--$o_4$) and objects $o_3$ and $o_4$ have missing temporal information. All trajectories start at time 0.

%% file: 4_real_time_compression.tex
\section{ONTRAC}

This section describes \textbf{ONTRAC} (\textbf{ON}line \textbf{TRA}jectory \textbf{C}ompression), a new framework for online map-matched trajectory compression. Its basic idea is building prediction models for trajectories and applying these models to suppress updates to a database. 

ONTRAC decomposes a trajectory $T$ into spatial and temporal components. The spatial information ($T.spatial$) is a sequence of road segments $<s_1, \ldots s_m>$ and the temporal information ($T.temporal$) is a sequence $<(d_1,t_1) \ldots $ $(d_m,t_m)>$ where $d_i$ is a distance and $t_i$ is a timestamp. Such a decomposition is motivated by the fact that spatial and temporal trajectory information have different properties, while the spatial information is discrete (road segments), the temporal is continuous (travel-times). 

Sections \ref{subsec::spatial_compression} and \ref{subsec::temporal_compression} describe ONTRAC's spatial and temporal compression strategies, respectively. Although the same general approach is applied for both cases --- learning a model to predict future updates based on past data --- the problem becomes more challenging when we move from the spatial to the temporal domain. Section \ref{subsec::querying} discusses how to query trajectory data compressed using ONTRAC.

\subsection{Spatial Compression}
\label{subsec::spatial_compression}

ONTRAC compresses the spatial information from trajectories using a Markov model learned during a training phase. The model is applied to predict the next update (i.e. road segment) for a given trajectory given its past updates. Compressed trajectories contain those segments that are not predicted by the model. This approach leads to a lossless compression, since any trajectory can be fully recovered as a combination of its compressed form and the model.

A $k$-order Markov trajectory model $\Psi$ for predicting the most likely next segment update $s$ for a given spatial trajectory $S$ is learned from training data $\mathcal{D}_t$: 

\begin{equation}
\Psi(S, \mathcal{D}_t) = \argmax_s P(s|S')
\label{equation::ppm_compression}
\end{equation}
where $S'$ is the longest prefix of $S$ in $\mathcal{D}_t$ such that $|S'| \leq k$ and $P(s|S')$ is the probability of $s$ to happen given $S'$ in $\mathcal{D}_t$. 

A \textit{Trie} data structure is applied as a representation for $\Psi$. Each node $nd$ in $\Psi$ corresponds to a sequence of road segments and has a vector $count$ that keeps track of the frequency of next segments and a variable $pred$ set to the most frequent next segment for the corresponding sequence. 

Algorithm \ref{alg:PPM_training} describes the training algorithm (\textbf{Spatial-training}) for spatial compression. It receives the road network $G$, the training trajectories $\mathcal{D}_t$, and the order $k$ as parameters and returns a $k$-order Markov model $\Psi$ for trajectory compression. For each trajectory $T$ in $\mathcal{D}_t$, the algorithm extracts its spatial information $S$ (line 3). A sliding window $W$ of size up to $k+1$ is moved along $S$ (lines 4 and 5) and the segments in $W$ are added to $\Psi$ in reverse order (lines 6-12). For every size $w \in [1,|W|-1]$ suffix of $W$, the next segment $W[w+1]$ has its count ($nd.count[W[w+1]]$) updated (line 9). The prediction $nd.pred$ associated to the node $nd$ is also updated if needed (line 10).

\begin{algorithm}[ht!]
\scriptsize
\caption{Spatial-training \label{alg:PPM_training}}
\begin{algorithmic}[1]
\REQUIRE Road network $G$, training data $\mathcal{D}_t$, order $k$
\ENSURE	Markov chain model $\Psi$
\STATE $\Psi \leftarrow$ empty tree
\FOR{Trajectory $T \in \mathcal{D}_t$}
\STATE $S \leftarrow T.spatial$
\FOR{$s \in [1,|S|]$}
\STATE Window $W \leftarrow S[\max(1,s-k):s]$
\STATE Node $nd \leftarrow \Psi.root$
\FOR{$w \in [|W|-1,1]$}
\STATE $nd \leftarrow nd[W[w]]$
\STATE $nd.count[W[w+1]] \leftarrow nd.count[W[w+1]] + 1$
\STATE $nd.pred \leftarrow \max_{p} nd.count[p]$
\ENDFOR
\ENDFOR
\ENDFOR
\end{algorithmic}
\end{algorithm}
\begin{algorithm}[ht!]
\scriptsize
\caption{Spatial-compression \label{alg:PPM_compression}}
\begin{algorithmic}[1]
\REQUIRE Trajectory $T$, model $\Psi$, order $k$
\ENSURE	Compressed spatial trajectory $S'$
\STATE $S \leftarrow T.spatial$
\STATE $S' \leftarrow <S[1]>$
\FOR{$s \in [2,|S|]$}
\STATE Window $W \leftarrow S[\max(1,s-k):s-1]$
\STATE $b \leftarrow |W|$
\STATE Node $nd \leftarrow \Psi.root[W[b]]$
\WHILE{$nd[W[b-1]] \neq \textbf{null}$ \textbf{AND} $b > 1$}
\STATE $b \leftarrow b - 1$
\STATE $nd \leftarrow nd[W[b]]$
\ENDWHILE
\IF{$S[s] \neq nd.pred$}
\STATE $S' \leftarrow S' \sqcup <S[s]>$
\ENDIF
\ENDFOR
\end{algorithmic}
\end{algorithm}

\begin{figure}[ht!]
\centering
\includegraphics[keepaspectratio, width=0.25\textwidth]{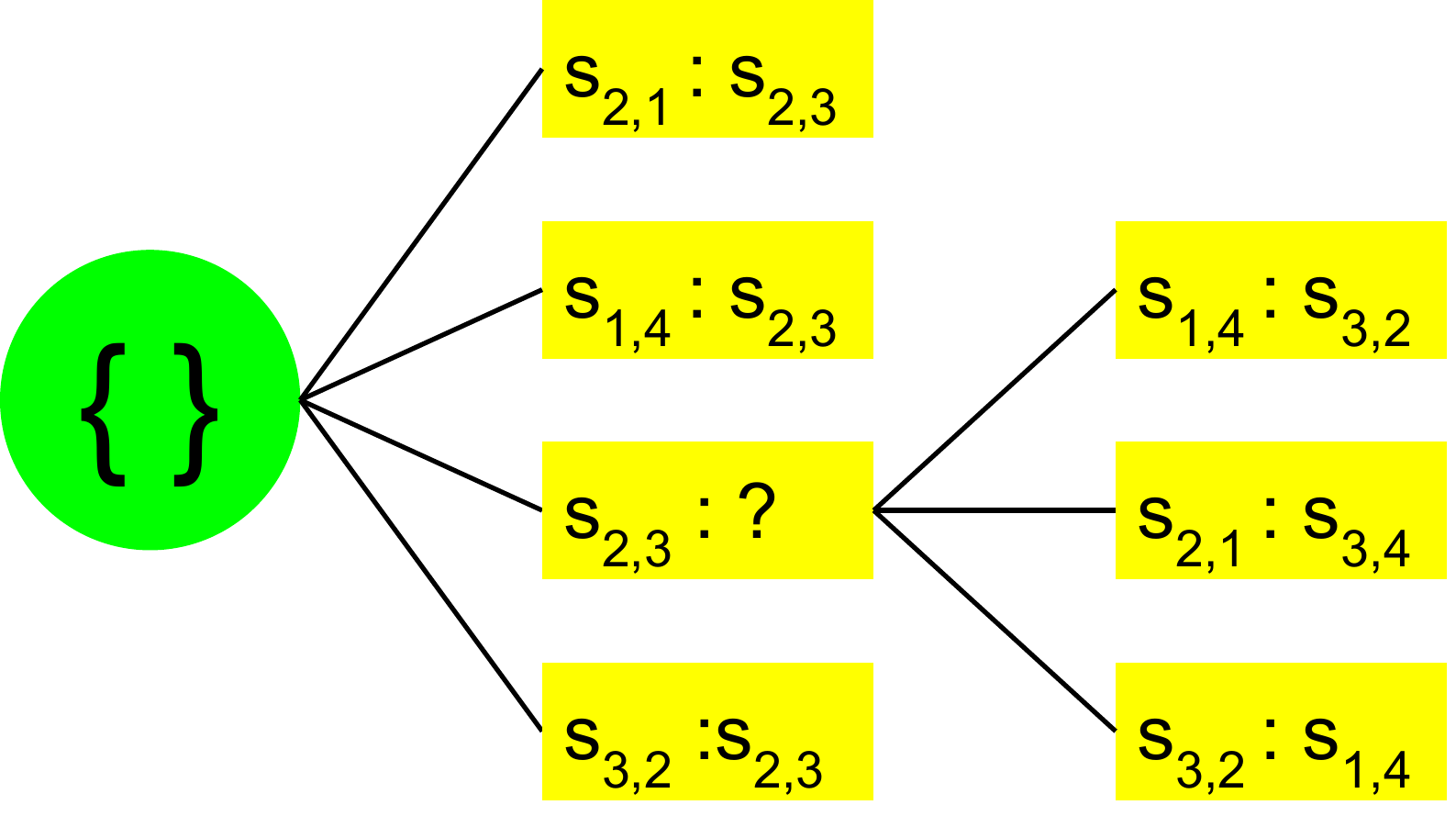}
\caption{Markov model learned from the training trajectories $\mathcal{D}_t=\{o_1,o_2,o_3\}$ in our running example ($k=2$).\label{fig:PPM_model}}
\end{figure}

The Markov model $\Psi$ learned in the training phase is applied for online compression of new trajectories, as described in Algorithm \ref{alg:PPM_compression} (\textbf{Spatial-compression}). Besides $\Psi$, the algorithm receives a trajectory $T$ and the order $k$ as parameters and returns the spatial information of $T$ in compressed form $S'$. Similar to Algorithm \ref{alg:PPM_training}, the compression algorithm extracts the spatial information $S$ from $T$ (line 1). The first segment ($S[1]$) is added to $S'$ (line 2) and, each further segment $S[s]$ ($2\leq s \leq |S|$) is compared against the prediction from $\Psi$ given a sliding window $W$ of size up to $k$ past segments (lines 3-14). Predictions are based on the maximum matching suffix of $W$ in $\Psi$. In case the prediction is incorrect, the new update $S[s]$ is added to $S'$ (line 12). We use the symbol `$\sqcup$' to denote sequence concatenation.

Figure \ref{fig:PPM_model} shows the Markov model $\Psi$ learned from the training trajectories $\mathcal{D}_t=\{o_1,o_2,o_3\}$ in our example (Table \ref{table:example_trajectories}) and for $k=2$. Each node is represented as a pair $s_i:s_j$ where $s_i$ is a segment for matching and $s_j$ is a prediction. For instance, in case the longest matching sequence of segments for a new update is $s_{2,1}$, the update is predicted as $s_{2,3}$. The compressed spatial information for the trajectory $o_4$ in our example is $<s_{1,2}, s_{2,1}>$, which gives a compression ratio of 2. Segments $s_{2,3}$ and $s_{3,4}$ were suppressed from the database, since they are predicted by the model.

Both the \textbf{spatial-training} and the \textbf{spatial-compression} algorithms are efficient in terms of space and time complexities. The size of the Markov model $\Psi$ is $O((deg^i)^k)$, where $deg^i$ is the maximum in-degree of a node in $E$, and \textbf{spatial-training} takes time $O(|\mathcal{D}_t||T|k)$, where $|T|$ is the maximum size of a trajectory. The time taken to compress a single trajectory using \textbf{spatial-compression} is $O(|T|k)$.

We define a new concept of \textit{spatial update block entropy} ($h_k$) as the average number of database updates required to represent a new trajectory update $s_{i+k+1}$ given that the previous $k$ updates $S' = <s_{i+1}, s_{i+2}, \ldots s_{i+k}>$ are known. As usual, we assume that trajectories are infinite and are generated from a stationary process (index $i$ is dropped).

\begin{align}
h_k &= \sum_{S \in \mathcal{D}} P(S)(1 - \delta(s_{k+1},\Psi(S', \mathcal{D}))) \nonumber \\
&= \sum_{S \in \mathcal{D}} P(S) - \sum_{S' \in \mathcal{D}} P(S'\sqcup \Psi(S',\mathcal{D})) \nonumber\\
&= 1 - \sum_{S' \in \mathcal{D}} P(S'\sqcup \Psi(S',\mathcal{D}))
\end{align}
where $S'=<s_1\ldots s_{k}>$, $S = S' \sqcup <s_{k+1}>$ and $\delta$ is the Kronecker delta function.

We apply the definition of $h_k$ to address an important question: \textit{How does the road network topology affect the spatial compression?} We assume a \textit{random-walk} trajectory model. In other words, given that the object is at a road segment $s_i$ it will choose any of the segments $s_j$ such that $(s_i,s_j) \in E$ uniformly at random. Similar to existing random-walk based algorithms \cite{page1999pagerank}, we add some connections to allow the trajectories to continue regardless of the segment they reach. Under this assumption, we can isolate the effect of the road network topology from the particular distribution of the trajectories in the data. 

\begin{thm}
Given a road network $G$ and a trajectory database $\mathcal{D}$ composed of random trajectories:
$$h_k = 1 - (\sum_{s \in V} \frac{\pi(s_i)}{deg^o(s_i)})$$
where $\pi(s_i)$ is the PageRank \cite{page1999pagerank} of $s_i$ and $deg^o(s_i)$ is the outdegree of $s_i$ in $G$.
\label{thm:network}
\end{thm}

Theorem \ref{thm:network} enables the comparison of two road networks (e.g. different cities), quantifying the hardness of compressing trajectories on them. For instance, it follows directly from this theorem that compressing trajectories in a fully connected road network is hard because the entropy $h_k$ for random trajectories approaches to 1. However, objects of relevance are not expected to move at random but rather follow mobility patterns \cite{gonzalez2008understanding}. Therefore, random trajectories should be more difficult to compress than actual ones. 

\subsection{Temporal Compression}
\label{subsec::temporal_compression}

The temporal compression algorithm applied by ONTRAC is also based on a prediction model learned using training data. However, instead of predicting road segments, here the goal is to predict a vehicle travel-times. Whenever a new update matches the one predicted by the model within an accepted error $\lambda$, the update is suppressed from the database. 

We start by defining a generative Gaussian model for travel-times (Section \ref{sec:gaussian_model}) based on some intuitive assumptions regarding vehicle trajectories. In Section \ref{sec:temporal_trajectory_inference}, we show how missing trajectory information can be inferred by fitting our Gaussian model via Maximum-likelihood Estimation, which leads to a convex Quadratic Programming formulation. This formulation is part of a learning algorithm (Section \ref{sec::temporal_training}) that estimates travel-time distributions for road segments based on Expectation-Maximization. Finally, we describe how the learned travel-time model is applied for trajectory compression in Section \ref{sec::temporal_compression}.

\begin{figure}[ht!]
\centering
\includegraphics[keepaspectratio, width=0.4\textwidth]{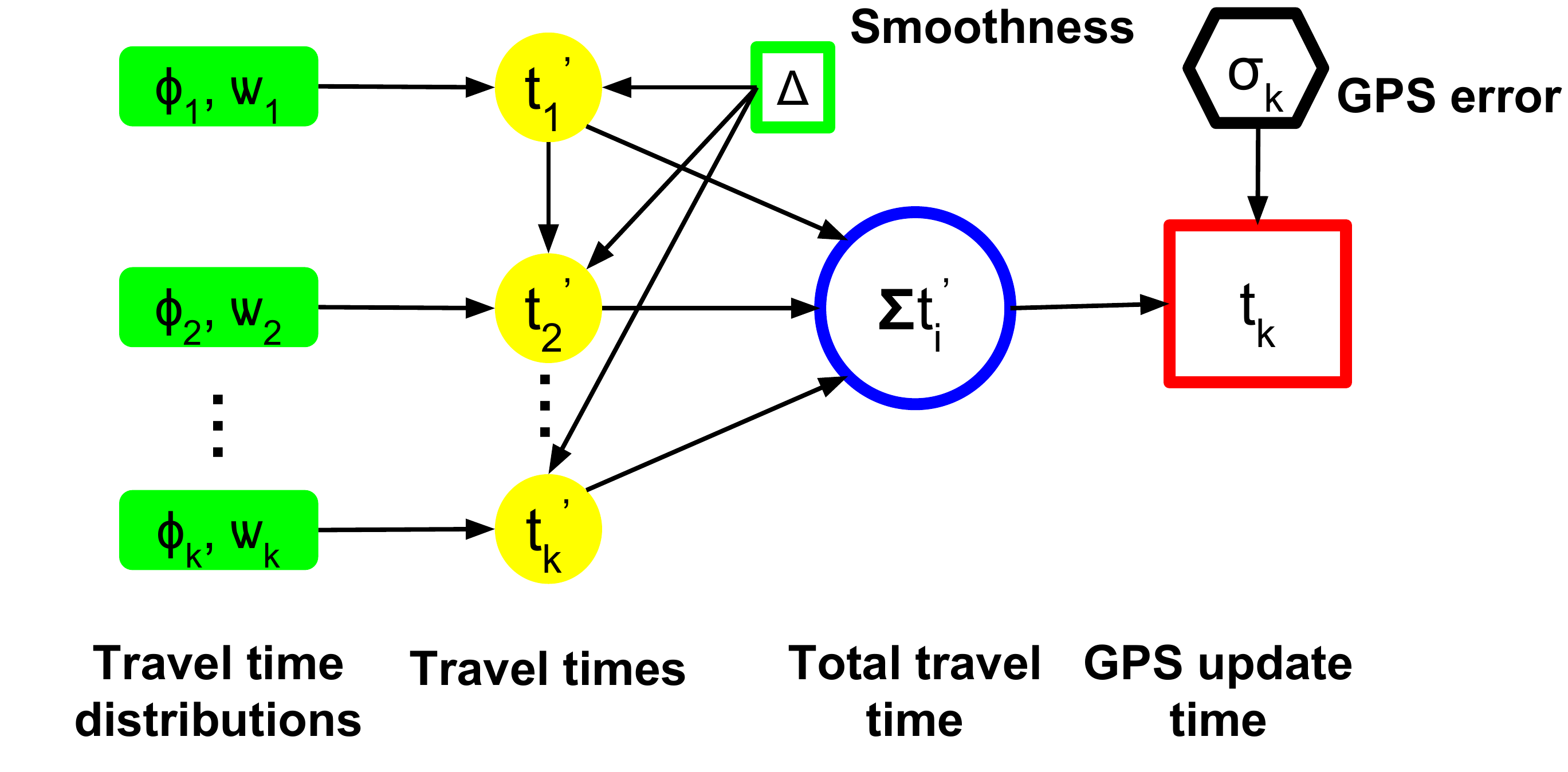}
\caption{Gaussian model for travel-times.\label{fig:gmtt_model}}
\end{figure}

\subsubsection{Generative Gaussian Model for Travel-times}
\label{sec:gaussian_model}

In order to predict temporal trajectory updates, ONTRAC assumes a generative Gaussian model for travel-times. Similar to existing approaches, the distributions involved in the model are Gaussians \cite{ide2011trajectory,gustafsson_wireless}. Such a model describes how the temporal information of GPS updates for a given trajectory is generated according to the expected travel-times for road segments and also taking into account the smoothness of vehicle speeds between segments. Figure \ref{fig:gmtt_model} provides a high-level description of ONTRAC travel-time model, which will be described in the remainder of this section.

As discussed in Section \ref{sec::online_trajectory_compression_and_inference}, temporal trajectory information is based on GPS updates, which are sparse and noisy. We use $t_i$ to represent total travel-times and $t_i'$ for relative times (i.e. $t_i'=t_i-t_{i-1}$). Segments and times are indexed by pairs $(j,i)$, where $j$ and $i$ are associated to the GPS update and segment/time, respectively. Given a sequence $<(s_1,t_1), (s_2,?), \ldots $ $(s_k,t_k)>$, with missing temporal information, $t_k = t_0 + \sum_{i=1}^k t_i' + \epsilon$ ($\epsilon \sim N(0,\sigma_k^2)$) , the GPS temporal error $\sigma_k$, associated to $t_k'$, can be estimated as:  

$$\sigma_k = \frac{\sigma^{\star}(t_k-t_{0})}{\sum_{i=1}^k |s_i|}$$
where $\sigma^*$ is the known GPS spatial error.

This error estimate follows from the speed formula $v=d/t$. Notice that GPS updates provide only the total time taken to traverse a sequence of segments but the travel-time $t_i'$ for each segment is missing. A simplistic approach is to assume that vehicles move at constant speed between updates (i.e. $t_j'= |s_j|(t_k-t_0)/\sum_{i=1}^k |s_i|$). However, due to the sparsity of GPS data (see Table \ref{table:data_stats}) such estimates are highly inaccurate. ONTRAC addresses this issue by learning travel-time distributions for segments from training data while enforcing a smoothness constraint on vehicle speeds.

ONTRAC assigns a travel-time distribution $N(\phi_i, \omega_i^2)$ to each segment $s_i \in E$. Vehicles traversing $s_i$ will draw their travel-time $t_i'$ from its distribution, as shown in Figure \ref{fig:gmtt_model}. Therefore, the total travel-time is such that:

$$\sum_{i=1}^k t_i' \sim N(\sum_{i=1}^k \phi_i,\sum_{i=1}^k \omega_i^2)$$

One important aspect of real trajectories is that physical constraints enforce speeds to be smooth (i.e. similar on nearby segments). We take smoothness into account in our model by adding a parameter $\Delta$, which controls variations in vehicle speeds through adjacent road segments as follows:

$$\frac{t_i'}{|s_i|} \sim N(\frac{t_{i-1}'}{|s_{i-1}|}, \Delta^2)$$

In the next section, we propose an algorithm for inferring travel-times given a sequence of GPS updates and model parameters. This inference step is applied by ONTRAC as means to predict temporal trajectories in the online compression. We use the notation ($\Phi,\Omega$) to denote the union of the parameters ($\phi_i,\omega_i$) for all the road segments in $G$.

\subsubsection{Temporal Trajectory Inference}
\label{sec:temporal_trajectory_inference}

Assuming that travel-times are generated according to our Gaussian model with known parameters (GPS error $\sigma^{\star}$, travel-time distributions $(\Phi,\Omega)$ and speed smoothness $\Delta$), ONTRAC estimates missing travel-times $t_i'$ for segments in a given trajectory from its GPS updates by means of Maximum-likelihood Estimation \cite{edward1972likelihood}. Here, we propose a Quadratic Programming (QP) formulation for maximizing the likelihood of a trajectory. This inference step can be performed efficiently because the resulting QP is convex \cite{boyd2004convex}.

Without loss of generality, we will assume that every GPS update covers $k$ road segments. Let $T = <(s_{1,1},t_{1,1}), (s_{1,2},?),$ $\ldots (s_{1,k},t_{1,k}), \ldots (s_{n,k},t_{n,k})>$ be a trajectory with $n$ GPS updates and some missing temporal information. The sequence $T'=$ $<t_{1,1}' \ldots t_{1,k}' \ldots t_{n,k}'>$ contains the estimated travel-times for $T$. Under the assumption of independence between the Gaussian trajectory model components, the likelihood of travel-times $T'$, $P(T')$, for temporal trajectory $T$ given the parameters $\sigma^{\star}$, $(\Phi,\Omega)$, and $\Delta$ is computed as:

\begin{scriptsize}
\begin{equation}
\underbrace{\prod_{i = (1,1)}^{(n,k)} P(t_i'|\phi_i,\omega_i) }_\text{travel time distribution} \overbrace{\prod_{i={(1,2)}}^{(n,k)} P(t_i'|t_{i-1}',\Delta)}^\text{smoothness} \underbrace{\prod_{j=1}^{n} P(\sum_{i=(j,1)}^{(j,k)} t_i' | \overline{t}_j,\sigma^{\star})}_\text{GPS update}
\label{eqn:traj_like}
\end{equation}
\end{scriptsize}
where $\overline{t}_j=t_{j,k}-t_{j-1,k}$ is the travel-time in between updates $(s_{j-1,k},t_{j-1,k})$ and $(s_{j,k},t_{j,k})$. 

Our goal is to identify the travel-times $T_{max}'$ that maximize $P(T')$. However, devising an analytical formulation for  $T_{max}'$ has shown to be a quite elusive task. Instead, ONTRAC makes use of the assumption that the several distributions involved in our model are Gaussians, which enables an expansion of Expression \ref{eqn:traj_like} as a Quadratic Programming (QP) instance, as shown in Theorem \ref{thm:qp_form}.

\begin{thm} The following Quadratic Programming formulation is the maximum-likelihood estimate for travel-times $T'$ for a trajectory $T$ assuming the trajectory Gaussian model.
$$T_{max}'=\arg\min_{\textbf{x}} \frac{1}{2} \textbf{x}^{T}Q\textbf{x}+\textbf{c}^T\textbf{x}$$ \textbf{s.t.} $\textbf{x} \geq 0$, where:
$$Q=Q_1+Q_2+Q_3, \textbf{c}=\textbf{c}_1+\textbf{c}_2$$
$$Q_1=\begin{pmatrix}
\frac{1}{\sigma_{1,1}^2}& 0 & \cdots & 0 \\
0 & \frac{1}{\sigma_{1,2}^2} & \cdots & 0 \\
\vdots  & \vdots  & \ddots & \vdots  \\
0 & 0 & \cdots & \frac{1}{\sigma_{n,k}^2}
\end{pmatrix}
$$
$$Q_2=\begin{pmatrix}
\frac{1}{\Delta^2|s_{1,1}|^2}& \frac{-1}{2\Delta^2|s_{1,1}||s_{1,2}|} & \cdots & 0 \\
\frac{-1}{2\Delta^2|s_{1,1}||s_{1,1}|}& \frac{1}{\Delta^2|s_{1,2}|^2}  & \cdots & 0 \\
 \vdots  & \vdots & \ddots & \vdots  \\
 0 & 0 & \cdots & \frac{1}{\Delta^2|s_{n,k}|^2}
\end{pmatrix}
$$
$$Q_3=\begin{pmatrix}
Q_3^1& 0 & \cdots & 0 \\
0 & Q_3^2  & \cdots & 0 \\
\vdots  & \vdots & \ddots & \vdots  \\
 0 & 0 & \cdots & Q_3^n
\end{pmatrix}
,Q_3^j=\begin{pmatrix}
\frac{1}{\sigma_j^2}& \cdots & \frac{1}{\sigma_j^2}\\
 \vdots & \ddots & \vdots  \\
\frac{1}{\sigma_j^2}& \cdots & \frac{1}{\sigma_j^2}
\end{pmatrix}
$$
$$\textbf{c}_1=-\begin{bmatrix} \frac{\phi_{1,1}}{\omega_{1,1}^2} & \cdots & \frac{\phi_{n,k}}{\omega_{n,k}^2}\end{bmatrix}$$
$$ \textbf{c}_2=-\begin{bmatrix} \textbf{c}_2^1 & \cdots & \textbf{c}_2^n\end{bmatrix}, \textbf{c}_2^j=\begin{bmatrix} \frac{\overline{t}_j}{\sigma_j^2}& \cdots & \frac{\overline{t}_j}{\sigma_j^2}\end{bmatrix}$$
\label{thm:qp_form}
\end{thm}

Theorem \ref{thm:qp_form} supports the use of the existing theory on \textit{Quadratic Optimization} to compute maximum-likelihood estimates for vehicle travel-times based on Equation \ref{eqn:traj_like}. However, general QP is NP-hard \cite{boyd2004convex} and an algorithm that is exponential on the number of road segments would prevent the application of our compression scheme to large trajectories. We show that our QP formulation is convex, and thus can be solved in polynomial time.

\begin{thm}
The Quadratic Programming problem described in Theorem \ref{thm:qp_form} is convex.
\label{thm:convex}
\end{thm}

We show how Theorems \ref{thm:qp_form} and \ref{thm:convex} are applied by ONTRAC using our running example (see Table \ref{table:example_trajectories}). Consider trajectory $o_3$ and assume $\phi_{3,2}=6$, $\phi_{2,3}=12$, $\phi_{1,4}=7$, $\omega_{3,2}=1$, $\omega_{2,3}=2$, $\omega_{1,4}=2$, $\Delta=2.5$ and $\sigma_{1,5}=3$. Then the matrix $Q$ and vector $\textbf{c}$ are computed as follows:

\begin{scriptsize}
$$Q=\begin{pmatrix}
1 & 0 & 0 \\
0 & .25  & 0 \\
0 & 0  & .25 \\
\end{pmatrix} + 
\begin{pmatrix}
.01 & -.005 & 0 \\
-.005 & .01  & -.005 \\
0 & -.005  & .01 \\
\end{pmatrix} + 
\begin{pmatrix}
.11 & .11 & .11 \\
.11 & .11  & .11 \\
.11 & .11  & .11 \\
\end{pmatrix}  
$$

$$
\textbf{c} = [-6, -3, -1.75] + [1.89, 1.89, 1.89]
$$
\end{scriptsize}

By giving the resulting QP instance to a solver (see Section \ref{sec:experiments}), we obtain maximum-likelihood travel-times $t_{3,2}=5.6$, $t_{2,3}=1.0$ and $t_{1,4}=5.3$. In the next section, we describe how these estimates are applied by ONTRAC's training algorithm for travel-time prediction.

\subsubsection{Training}
\label{sec::temporal_training}

ONTRAC's temporal trajectory compression model consists of the average travel-time distributions for segments $(\Phi,\Omega)$. In order to learn the parameters of these distributions, we apply our QP formulation as part of an Expectation-Maximization (EM) algorithm \cite{dempster1977maximum} that iterates over the set of training trajectories until convergence.

\begin{algorithm}[ht!]
\scriptsize
\caption{Temporal-training \label{alg:EMKF_training}}
\begin{algorithmic}[1]
\REQUIRE Road network $G(V,E)$, training data $\mathcal{D}_t$, GPS error $\sigma^{\star}$, smoothness $\Delta$, initial model $(\Phi^0,\Omega^0)$ iterations $q$
\ENSURE	travel-time distributions $(\Phi^q,\Omega^q)$
\FOR{$\ell \in [1,q]$ iterations}
\STATE $\Phi^{\ell}\leftarrow \emptyset$
\STATE $\Omega^{\ell}\leftarrow \emptyset$
\FOR{Trajectory $T_i \in \mathcal{D}_t$}
\STATE $T'\leftarrow \argmax_T P(T|\Phi^{\ell-1}, \Omega^{\ell-1}, \Delta, \sigma^{\star})$
\ENDFOR
\FOR{Segment $s_i \in V$}
\STATE $\mathcal{T}_i \leftarrow \{T \in \mathcal{D}_t| s_i \in T\}$
\STATE $\phi_i \leftarrow \frac{1}{|\mathcal{T}_i|} \sum_{T \in \mathcal{T}_i} t_i'$
\STATE $\omega_i \leftarrow \sqrt{\frac{1}{|\mathcal{T}_i|} \sum_{T \in \mathcal{T}_i} (t_i'-\phi_i)^2}$
\STATE $\Phi^{\ell}\leftarrow \Phi^{\ell}\cup \phi_i$, $\Omega^{\ell}\leftarrow \Omega^{\ell}\cup \omega_i$; 
\ENDFOR
\ENDFOR
\end{algorithmic}
\end{algorithm}

ONTRAC \textbf{Temporal-training} algorithm (Algorithm \ref{alg:EMKF_training})  receives the road network $G$, the training data $\mathcal{D}_t$, the GPS error $\sigma^{\star}$, the smoothness parameter $\Delta$, an initial model $(\Phi^0, \Omega^0)$, and the number of iterations $q$. The algorithm returns a maximum-likelihood model $(\Phi^q,\Omega^q)$ given the data and other parameters.  We formulate our EM scheme as follows. Training trajectories $\mathcal{D}_t$ are observed data, travel-time distributions $(\Phi,\Omega)$ are parameters and travel-times $t_i'$ for each trajectory are latent variables. The algorithm starts with an initial setting for the model $(\Phi^0.\Omega^0)$. In the \textit{expectation step (e-step)}, expected values for the latent variables given the parameters are computed using Quadratic Programming (line 5). Next, in the \textit{maximization (m-step)}, the model $(\Phi^{\ell},\Omega^{\ell})$ is updated to maximize the likelihood of the data given the latent variables computed in the \textit{e-step} (lines 7-12). After a finite number of iterations, the algorithm converges to a local optima, as shown in Theorem \ref{thm::convergence}.

\begin{thm}
The \textbf{Temporal-training} estimate for the model $(\Phi,\Omega)$ converges to a local optima.
\label{thm::convergence}
\end{thm}

\textbf{Temporal-training} runs in $O(q(|\mathcal{D}_t||T|^3 + |V|))$ time, where $q$ is the number of iterations, $|T|$ is the maximum trajectory length and $V$ is the set of segments. Solving a QP instance takes $O(|T|^3)$ time and the algorithm solves $n$ QP instances in the \textit{e-step}, updating the travel-time distribution parameters for each segment $s_i \in V$ in the \textit{m-step}.

\subsubsection{Compression}
\label{sec::temporal_compression}

ONTRAC applies the travel-time model $(\Phi,\Omega)$ learned by the \textbf{Temporal-training} algorithm to predict travel-times.  Updates are suppressed if the model predictions are within a user-defined accepted time error $\lambda$ from the actual travel-times. Therefore, ONTRAC's temporal compression is lossy.

Update travel-times are computed by fusing the model ($\Phi,\Omega$) and the GPS update information. Given our Gaussian assumptions, the problem consists of fusing the means of two Gaussians with known variance as a weighted mean:
\begin{equation}
\frac{\hat{t}\sigma_i^2+\overline{t}_i\hat{w}}{\hat{w}+\sigma_i^2}
\end{equation}
where $\hat{t}$ is the model prediction with variance $\hat{w}$ and $\overline{t}_i$ is the GPS time update with variance $\sigma_i^2$.

For missing updates, ONTRAC assumes that travel-times $t_j$ are proportional to the estimates from the model $(\Phi,\Omega)$:
\begin{equation}
\frac{(\hat{t}\sigma_i^2+\overline{t}_i\hat{w})\phi_j}{(\hat{w}+\sigma_i^2)\hat{t}}
\label{equation::temporal_proportion}
\end{equation}

\begin{algorithm}[ht!]
\scriptsize
\caption{Temporal-compression \label{alg:EMKF_compression}}
\begin{algorithmic}[1]
\REQUIRE Trajectory $T$, travel-time model $(\Phi,\Omega)$, GPS error $\sigma^{\star}$, accepted error $\lambda$
\ENSURE	Compressed temporal trajectory $D$
\STATE $D \leftarrow <>$; $t^{\star} \leftarrow t_0$; $d \leftarrow 0$
\STATE $\tau \leftarrow t_0$; $\hat{t} \leftarrow 0$; $\hat{w} \leftarrow 0$
\FOR{updates $i \in [1,|T|]$}
\STATE $\hat{t} \leftarrow \hat{t} + \phi_i$; $\hat{w} \leftarrow \hat{w} + \omega_i^2$; $d \leftarrow d + |s_i|$
\IF{segment $t_i$ is non-missing}
\STATE $t^{\star} \leftarrow t^{\star} + \frac{\hat{t}\sigma_i^2+\overline{t}_i\hat{w}}{\hat{w}+\sigma_i^2}$
\STATE $\tau \leftarrow \tau + \hat{t}$
\IF{error $|\tau-t^{\star}| > \lambda$}
\STATE Add update $D \leftarrow D \sqcup <(d,t^{\star})>$; $\tau \leftarrow t^{\star}$ 
\ENDIF
\STATE $\hat{w} \leftarrow 0$; $\hat{t} \leftarrow 0$
\ENDIF
\ENDFOR
\end{algorithmic}
\end{algorithm}

\begin{figure}[ht!]
\centering
\includegraphics[keepaspectratio, width=0.4\textwidth]{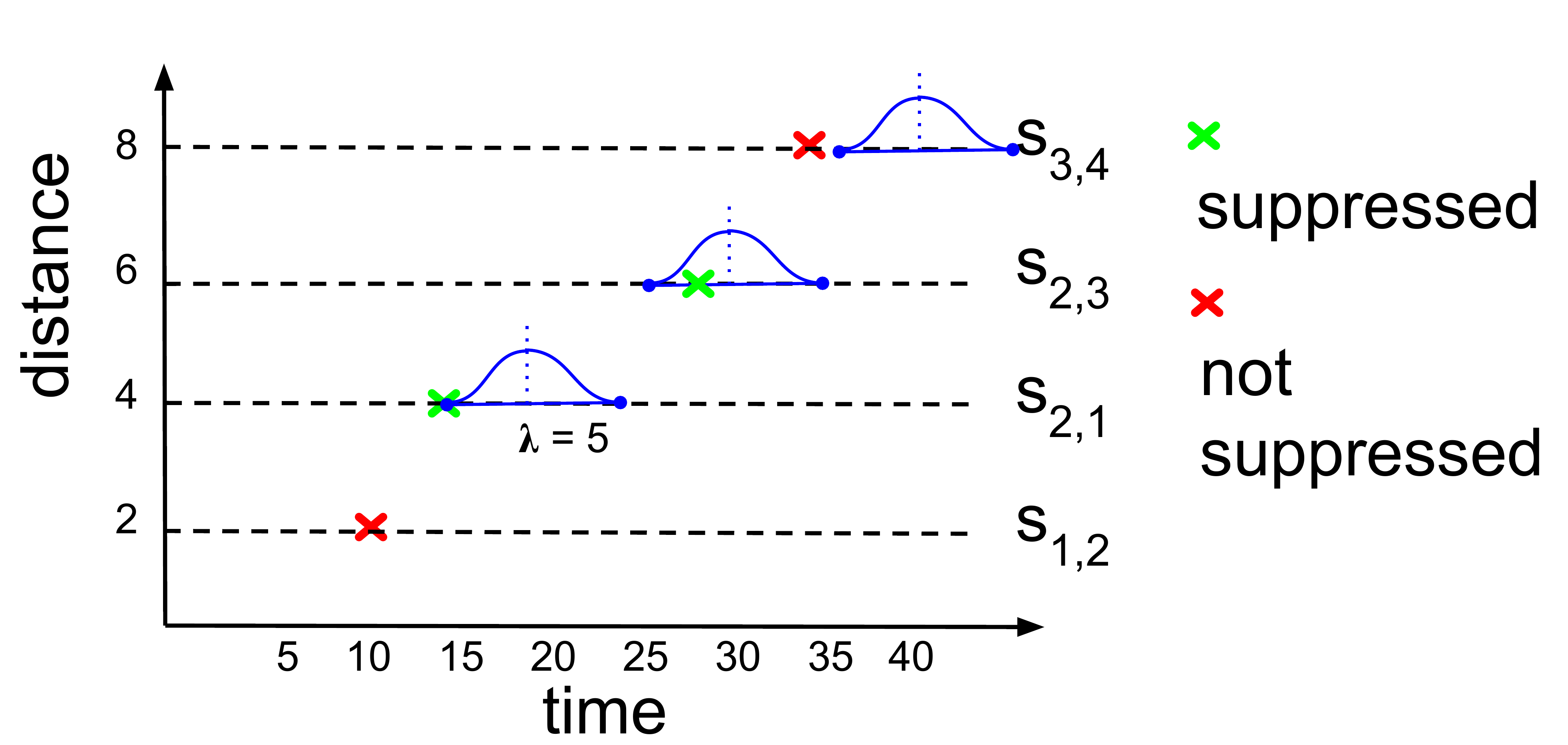}
\caption{Example of temporal compression for trajectory $o_4$ and $\lambda=5$. \label{fig:temporal_compression_example}}
\end{figure}

As a consequence, by keeping the model and the compressed version of the non-missing updates, ONTRAC guarantees the recovery of the missing ones within error $\lambda$.

Algorithm \ref{alg:EMKF_compression} describes ONTRAC's temporal compression algorithm. It receives a trajectory $T$, the prediction model ($\Phi,\Omega$), the GPS error $\sigma^{\star}$, and the time error $\lambda$ as parameters and returns a compressed version $D$ of $T$, which is initialized as empty (line 2). Also, both the actual total travel-time $t^{*}$ and the model estimate $\tau$ are initialized with the starting time of the trajectory (lines 2-3). For a non-missing update in $T$ the model travel-time prediction $\hat{t}$ is fused with the GPS time $\overline{t}_i$ in order to compute the actual travel-time $t^{\star}$ (line 6). If the predicted total time $\tau$ is within $\lambda$-error from the actual total time $t^{\star}$, the update is suppressed from $D$. Otherwise, the update is added to $D$ and the predicted total travel-time $\tau$ is corrected to the actual one (lines 9-10).

Figure \ref{fig:temporal_compression_example} illustrates ONTRAC's temporal compression for trajectory $o_4$, time error $\lambda=5$ and assuming time $\phi_{i,j}= 10$ for any segment $s_{i,j}$ in our running example. Two updates ,$(s_{2,1},15)$ and $(s_{2,3},?)$, are suppressed based on the prediction model. We assume the GPS update and fused times are the same in this example for the sake of simplicity.

The time complexity of \textbf{Temporal-compression} for a trajectory $T$ is linear with the size of the trajectory ($O(|T|)$). Moreover, the algorithm guarantees the upper bound on the time error $\lambda$ for any update in the compressed database.

\subsection{Querying}
\label{subsec::querying}

We describe how ONTRAC supports a type of query that returns the location of an object at a particular time. 

\begin{mydef}
\textbf{where($o_i,t$) query:} Given a database $\mathcal{D}$, an object $o_i$, and a time $t$, return the road segment $s_j$ where $o_i$ is located at time $t$ (i.e. $<s_j,t> \in T_i$).
\label{def::where_query}
\end{mydef}

\begin{algorithm}[ht!]
\scriptsize
\caption{Partial-decompression \label{alg::partial_decompression}}
\begin{algorithmic}[1]
\REQUIRE Compressed trajectory $T$, models: $\Psi$, $\Phi$, order $k$, time $t$
\ENSURE	Partially decompressed trajectory $\hat{T}$
\STATE Build context $q \leftarrow i | t_i \prec t_r$; $p \leftarrow i | t_i \prec t_{q-k}$
\WHILE {possible predictions $|\{\Psi'(\mathcal{T}[<s_p \to s_q>])\}| > 1$}
\STATE Move backwards $p \leftarrow i | t_i \prec t_p$ 
\ENDWHILE
\STATE Build sub-trajectory $\hat{T} \leftarrow <s_p \ldots s_q>$
\WHILE{$t_q < t$}
\STATE Reconstruct $\hat{T} \leftarrow \hat{T} \sqcup [\Psi,\Phi](\hat{T})$; $q \leftarrow |\hat{T}|$
\ENDWHILE
\end{algorithmic}
\end{algorithm}

For instance, $where(o_4,10)=s_{1,2}$. Notice that, different from the full data, ONTRAC's compressed trajectories cannot be queried without information from the compression model. In particular, because spatial information is compressed based on prefix matching (see Equation \ref{equation::ppm_compression}), recovering a trajectory requires contextual information that might also have been compressed. A simplistic approach to address this problem is performing a full reconstruction of a compressed trajectory $T$ at query time, which might incur a significant overhead on querying performance for long trajectories. Instead, ONTRAC supports the recovery of a partial decompression $\hat{T}$, which is a sub-trajectory of the decompressed version of $T$ that covers a query time $t$. 

The idea behind partial trajectory decompression is building a trajectory context that goes far enough backwards in time to make the reconstruction unique. We use the notation $\mathcal{T}[<s_p \to s_q>]$ to represent all the possible sub-trajectories in the form $<s_p, s_{p+1} \ldots s_q>$ where $s_{q}=\Psi(<s_p \ldots s_{q-1}>)$. Moreover, $\Psi'(\mathcal{T}[s_p \to s_q])$ gives the corresponding set of next segment predictions for sub-trajectories $\mathcal{T}[<s_p \to s_1>]$ in $\Psi$. Once there is a single next segment prediction, the trajectory can be uniquely reconstructed. 

Algorithm \ref{alg::partial_decompression} shows the partial trajectory decompression algorithm. It receives a compressed trajectory $T$, the spatial ($\Psi$) and temporal ($\Phi$) compression models, and the query time $t$. As a result, it returns $\hat{T}$ which is a partial decompression of $T$. The algorithm starts building a context $<s_p,s_q>$ that precedes the time $t$ and is long enough to cover a prefix of order $k$ (line 1). The segment $s_p$ is moved backwards in $T$ until it leads to a unique segment prediction (lines 2-4). The trajectory is then decompressed using the spatial and temporal model ($[\Psi,\Phi]$) starting from $s_p$ (lines 5-8). We use the symbol '$\prec$' to denote 'the latest before'.

For instance, consider the compressed trajectory\\ $<(s_{0,1},5), (s_{1,2},8), (s_{2,1},25), (s_{2,3},30)>$, a time $t=31$ for which a query requires a partial reconstruction, and the spatial prediction model $\Psi$ from Figure \ref{fig:PPM_model} with order $k = 2$. The algorithm will build a context $<s_{1,2} \to s_{2,3}>$ and compute the possible partial reconstructions $\Psi'(\mathcal{T}[s_{1,2}\to s_{2,3}]) = \{s_{3,4}\}$. Since this reconstruction is unique, the trajectory can be partially reconstructed using $<s_{1,2} \to s_{2,1}>$. Otherwise, it would move backwards along the compressed trajectory, adding $s_{0,1}$ to the context.

The \textbf{Partial-decompression} algorithm has  time complexity $O(|T|^2(deg^o)k)$, where $|T|$ is the maximum size of a decompressed trajectory, $deg^o$ is the maximum degree of a segment in $E$, and $k$ is the order of the spatial compression model $\Psi$. This is the same complexity of the querying procedure using the partial decompression scheme.

%% file: 5_experimental_results.tex
\section{Experimental Results}
\label{sec:experiments}

We evaluate ONTRAC online trajectory compression using real-world and synthetic datasets. Sections \ref{subsec::exp_spatial_compression} and \ref{subsec::exp_temporal_compression} show that our approach achieves high compression ratios, outperforming the baseline, in terms of spatial and temporal compression, respectively. Section \ref{subsec::exp_querying_storage}, demonstrates how ONTRAC improves database scalability in trajectory data management with a low overhead over the query time.

\begin{figure*}[ht!]
\centering
\subfloat[SF \label{fig::spatial_comp_ratio_sf}]{
\includegraphics[width=0.2\textwidth]{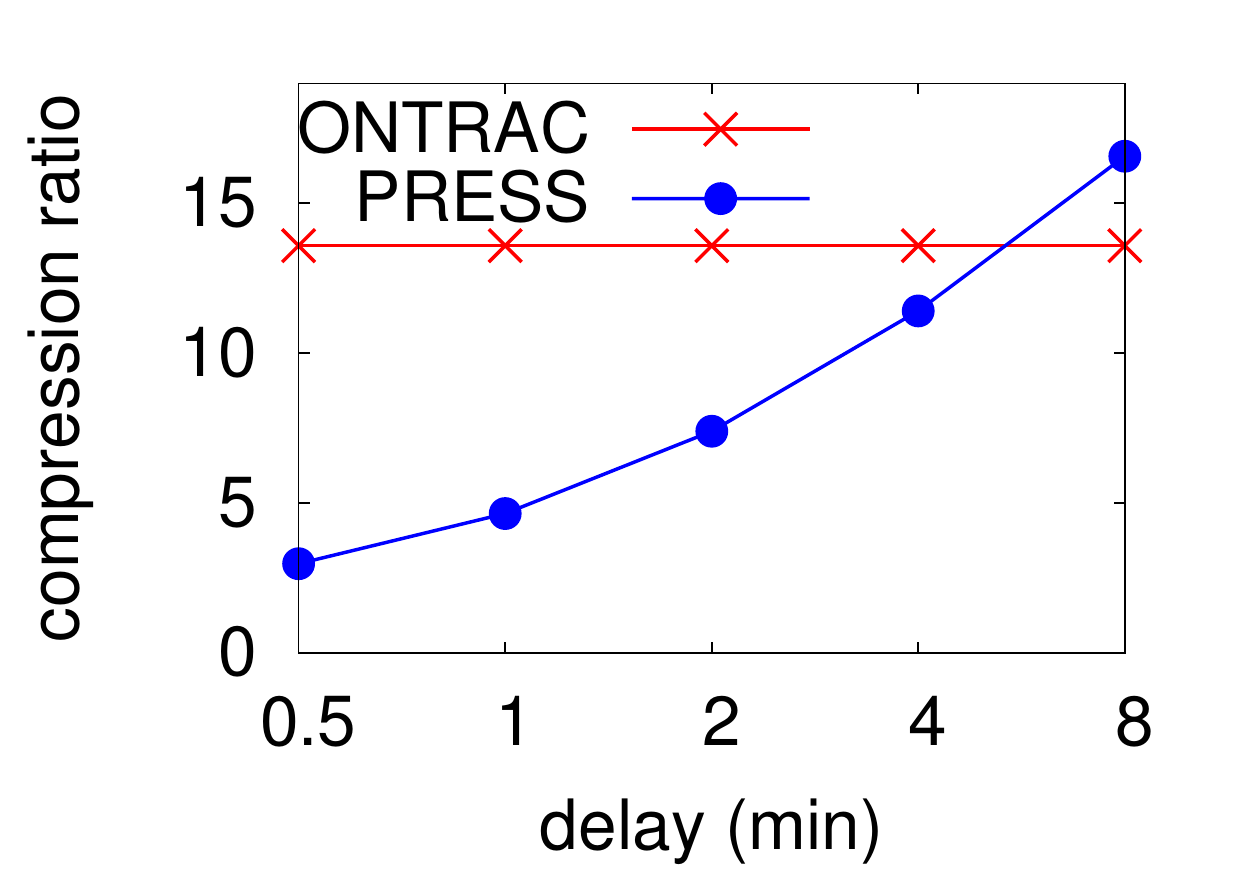}
}
\subfloat[Beijing \label{fig::spatial_comp_ratio_beijing}]{
\includegraphics[width=0.2\textwidth]{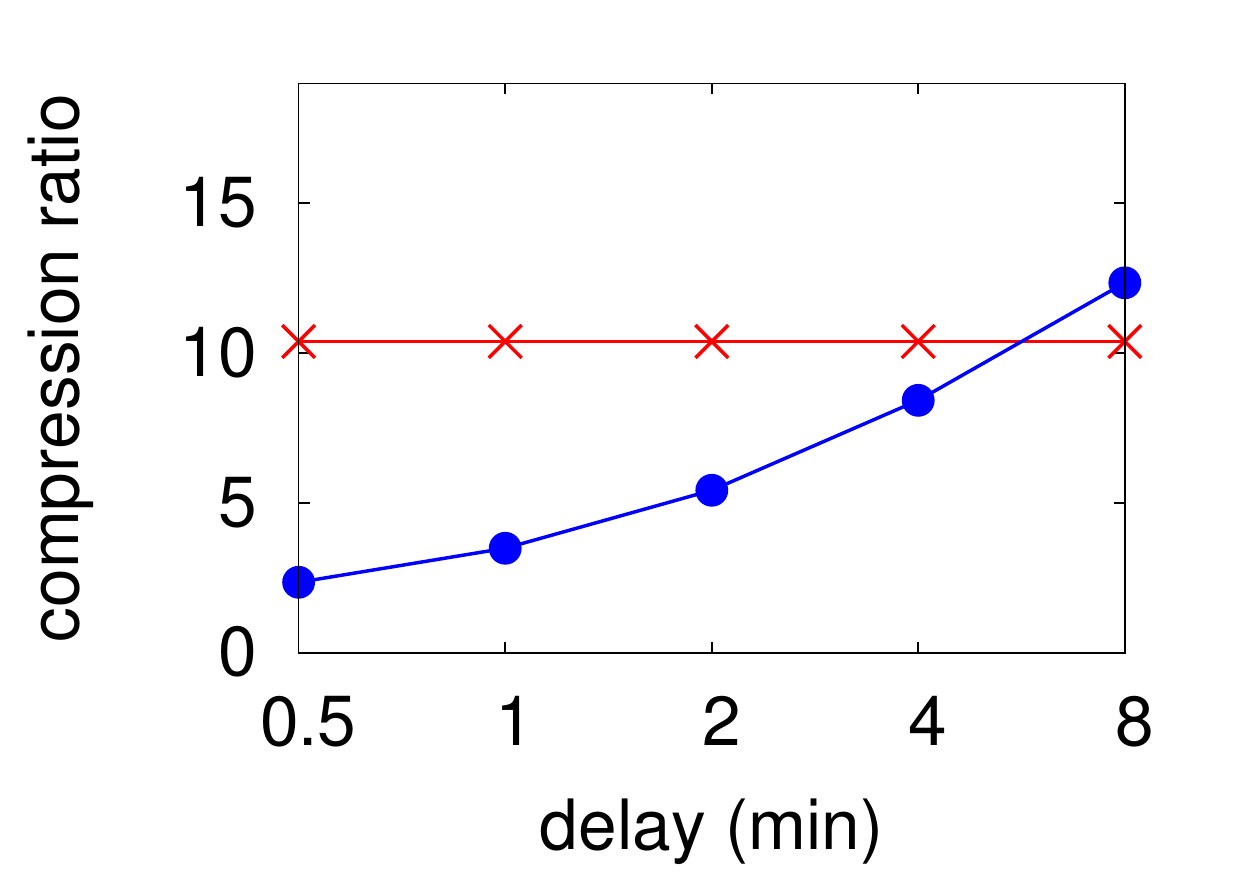}
}
\subfloat[SF \label{fig::spatial_comp_time_sf}]{
\includegraphics[width=0.2\textwidth]{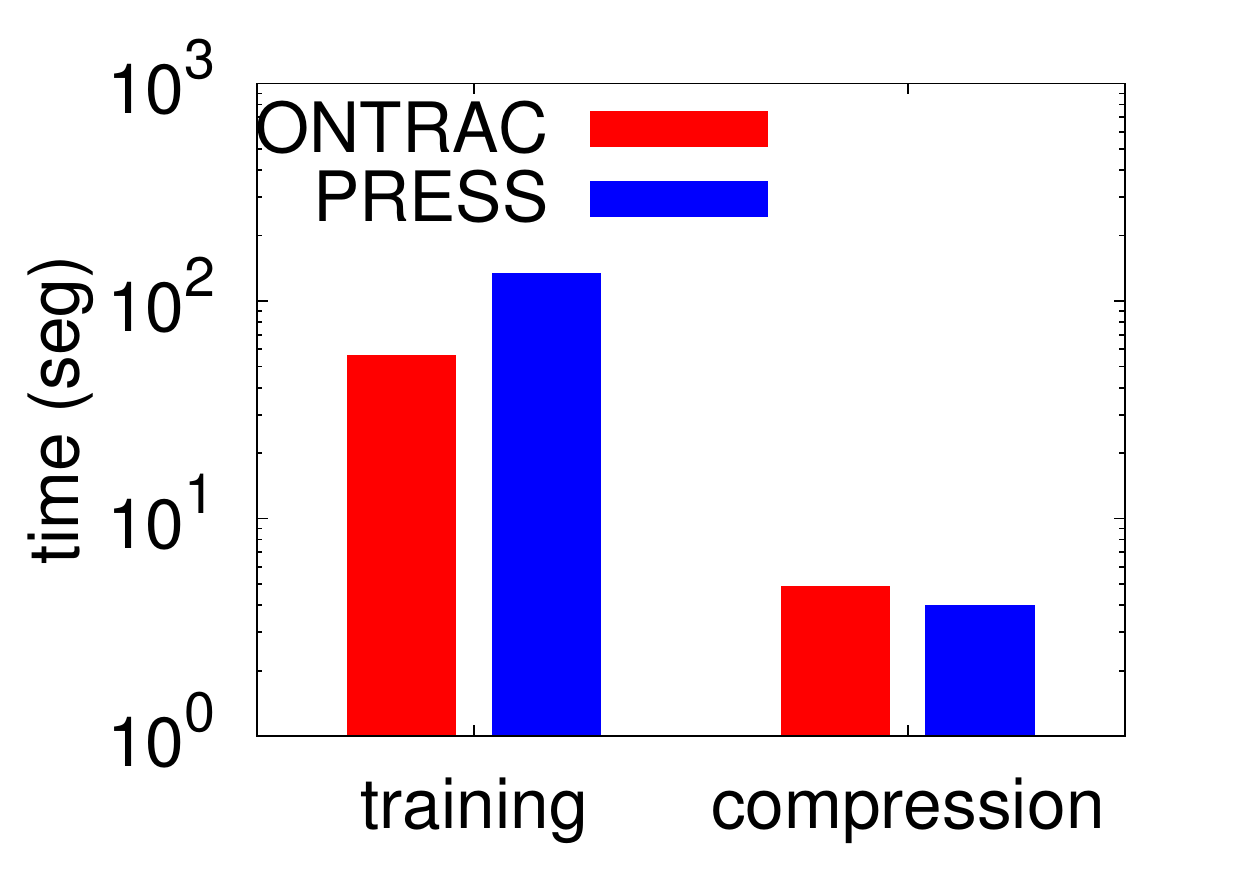}
}
\subfloat[Beijing \label{fig::spatial_comp_time_beijing}]{
\includegraphics[width=0.2\textwidth]{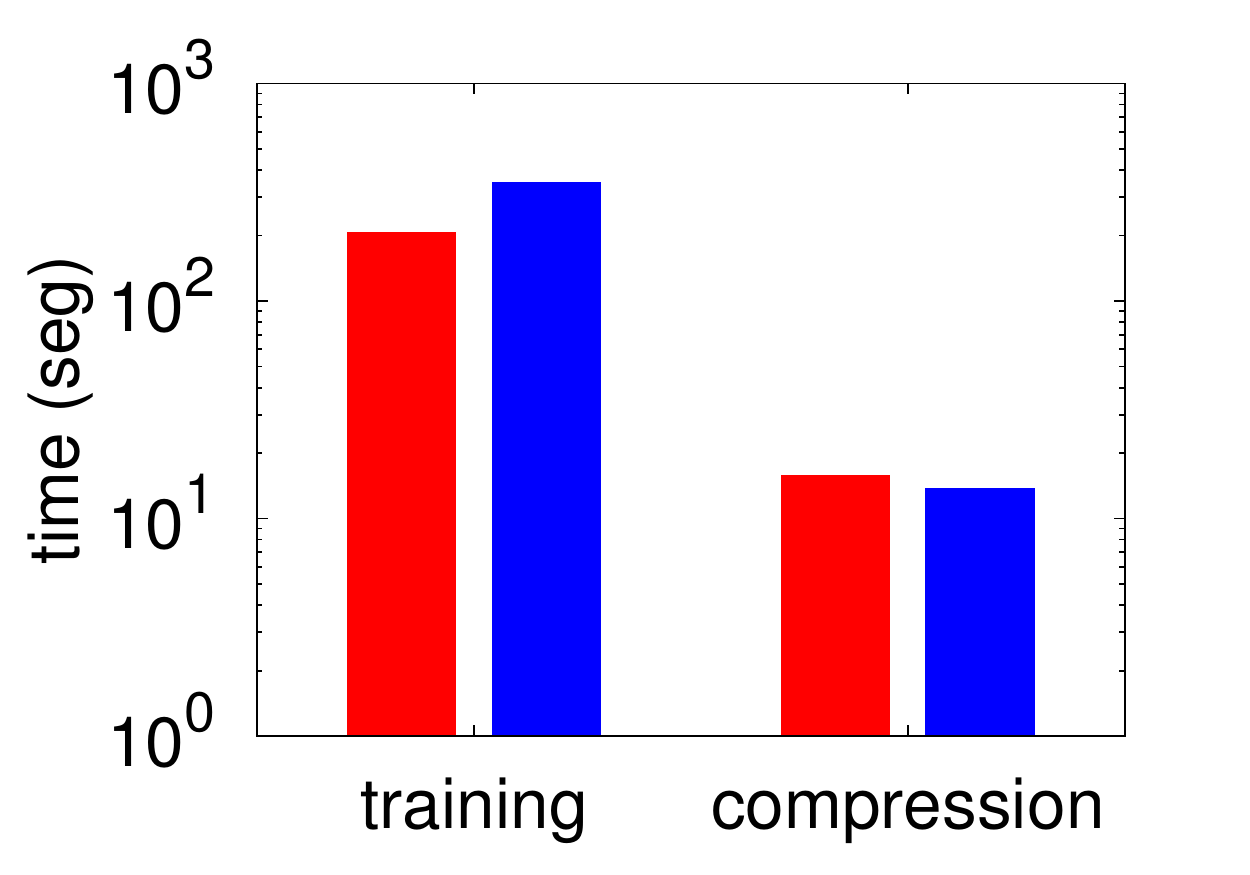}
}
\caption{Spatial compression ratio (a,b), training and compression times (c,d) for ONTRAC and PRESS using the SF (a,c) and Beijing (b,d) datasets. As a fully online scheme, ONTRAC achieves higher compression ratios than PRESS, which is not online, even when delays as long as 4 minutes are accepted. Also, ONTRAC achieves similar or better performance than PRESS in terms of training and compression time for most of the datasets. \label{fig::spatial_compression_results}}
\end{figure*}

Two datasets combining GPS taxi traces and road network data from \textit{OpenStreetMap} are applied in the experimental evaluation of ONTRAC. The first one is the \textit{San Francisco Cab} dataset (\textit{SF}) \cite{piorkowski2009parsimonious}, which consists of GPS traces from 500 cabs over 30 days in San Francisco. The second dataset (\textit{Beijing}) \cite{yuan2011driving} contains GPS traces generated by 10,357 cabs during a week-long period in Beijing. Table \ref{table:data_stats} shows the number of vertices and edges, the number of GPS updates, and the sampling rate (average time between two consecutive GPS updates) for the datasets.

\textbf{Baseline:} We extend PRESS \cite{pvldbSongSZZ14} to the online compression setting as a baseline for ONTRAC. A maximum delay is added as a parameter for both spatial and temporal compression. Updates can be compressed as in batch mode within the time delay and once the delay is reached, the trajectory database is updated. For temporal compression, we considered the maximum delay to be equal to $\lambda$. Moreover, we replace PRESS' Huffman codes generated for spatial trajectories by integer identifiers in order to add database support for PRESS compressed data.  

\textbf{Implementation:} The algorithms are implemented in C++ and are publicly available\footnote{\url{https://github.com/arleilps/traj-comp}}. Map-matching is performed using our implementation of a state-of-the-art algorithm \cite{6799840}. Querying and storage are implemented on top of \textit{PostgreSQL}\footnote{\url{http://www.postgresql.org/}} and its geographic extension \textit{PostGIS}\footnote{\url{http://postgis.net/}}. Quadratic programming is implemented using \textit{CPLEX}\footnote{\url{http://www-01.ibm.com/software/commerce/optimization/cplex-optimizer/}}. The training step of our temporal compression algorithm (Algorithm \ref{alg:EMKF_training}) was parallelized using \textit{pthreads}.

\textbf{Experimental Setting:} Experiments were run on a 2.67 GHz 8-core Intel i7 with 196GB RAM. We set the GPS error $\sigma$ to $5$ meters by default and the smoothness parameter $\Delta$ was estimated from the speed variance in the GPS updates. 

\subsection{Spatial Compression}
\label{subsec::exp_spatial_compression}

\begin{table}[t!]
\centering
\scriptsize
\begin{tabular}{| c | c | c | c | c |}
\hline
\textbf{name}& \textbf{\#vertices} & \textbf{\#edges} & \textbf{\#updates}& \textbf{samp. rate} \\
\hline
\textbf{SF} & 78,847 & 181,598 & 7.7M& 2.27m\\
\hline
\textbf{Beijing} & 461,710 & 904,144 & 12.6M& 6.86m\\
\hline
\end{tabular}
\caption{Dataset statistics. \label{table:data_stats}}
\end{table}

ONTRAC's spatial compression is evaluated in terms of compression ratio (in number of database updates), compression time and training time in Figure \ref{fig::spatial_compression_results}. We compare our approach against PRESS when different delays (in minutes) are allowed. For each dataset, 90\% of the trajectories are used for training and 10\% are used for compression.

Figures \ref{fig::spatial_comp_ratio_sf} and \ref{fig::spatial_comp_ratio_beijing} show the compression results of ONTRAC and PRESS with varying delays. ONTRAC achieves compression ratios of 13.6 and 10.4 for  \textit{SF} and \textit{Beijing}, respectively. Even when delays are up to 4 minutes, ONTRAC still achieves better compression results than the baseline. 

In Figures \ref{fig::spatial_comp_time_sf} and \ref{fig::spatial_comp_time_beijing}, we evaluate our approach in terms of training and compression time. Results show that ONTRAC achieves similar or better performance than the baseline approach for most of the datasets, being able to learn its compression model using up to 10 million updates in a time in the order of minutes. Moreover, ONTRAC can compress up to a million trajectories in time in the order of seconds.

\begin{figure*}[ht!]
\centering
\subfloat[SF \label{fig::temporal_comp_ratio_sf}]{
\includegraphics[width=0.2\textwidth]{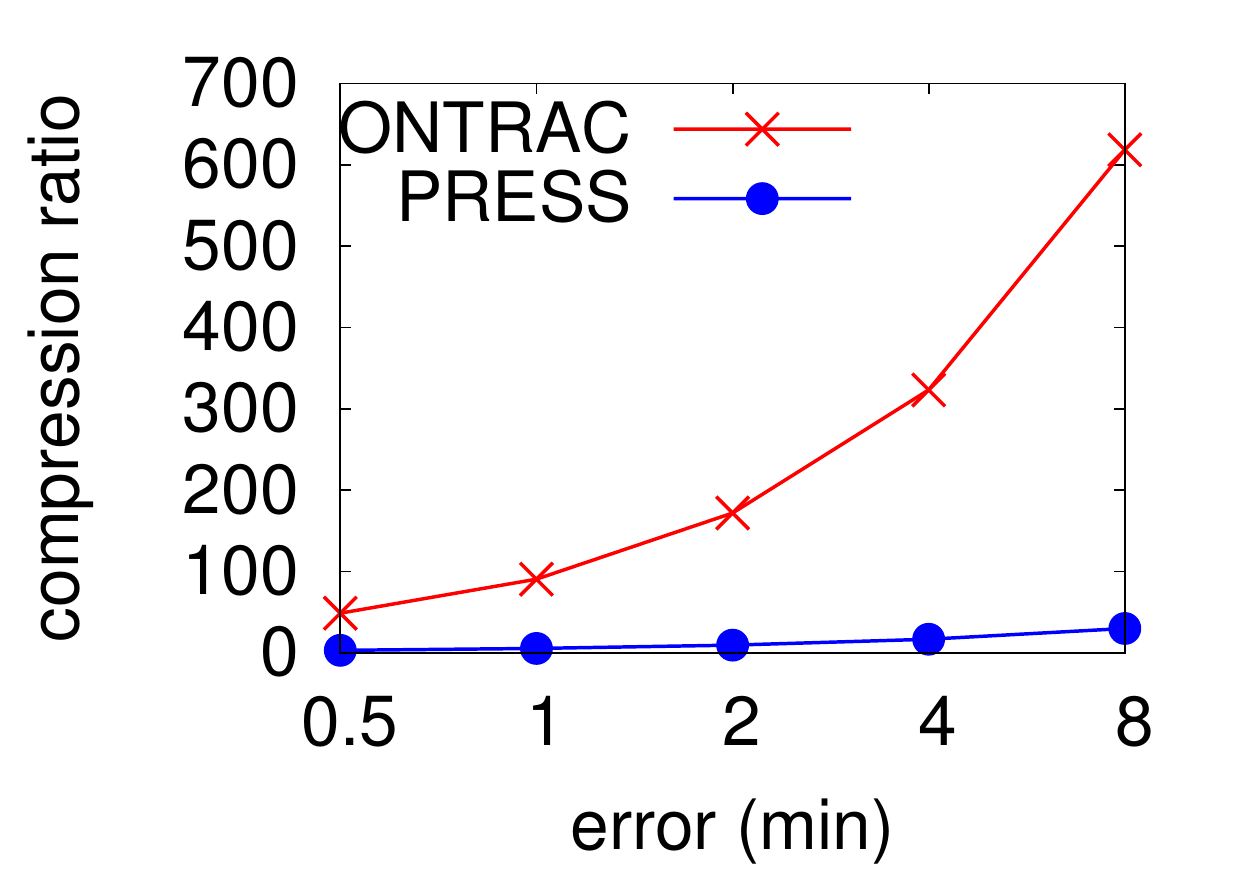}
}
\subfloat[Beijing \label{fig::temporal_comp_ratio_beijing}]{
\includegraphics[width=0.2\textwidth]{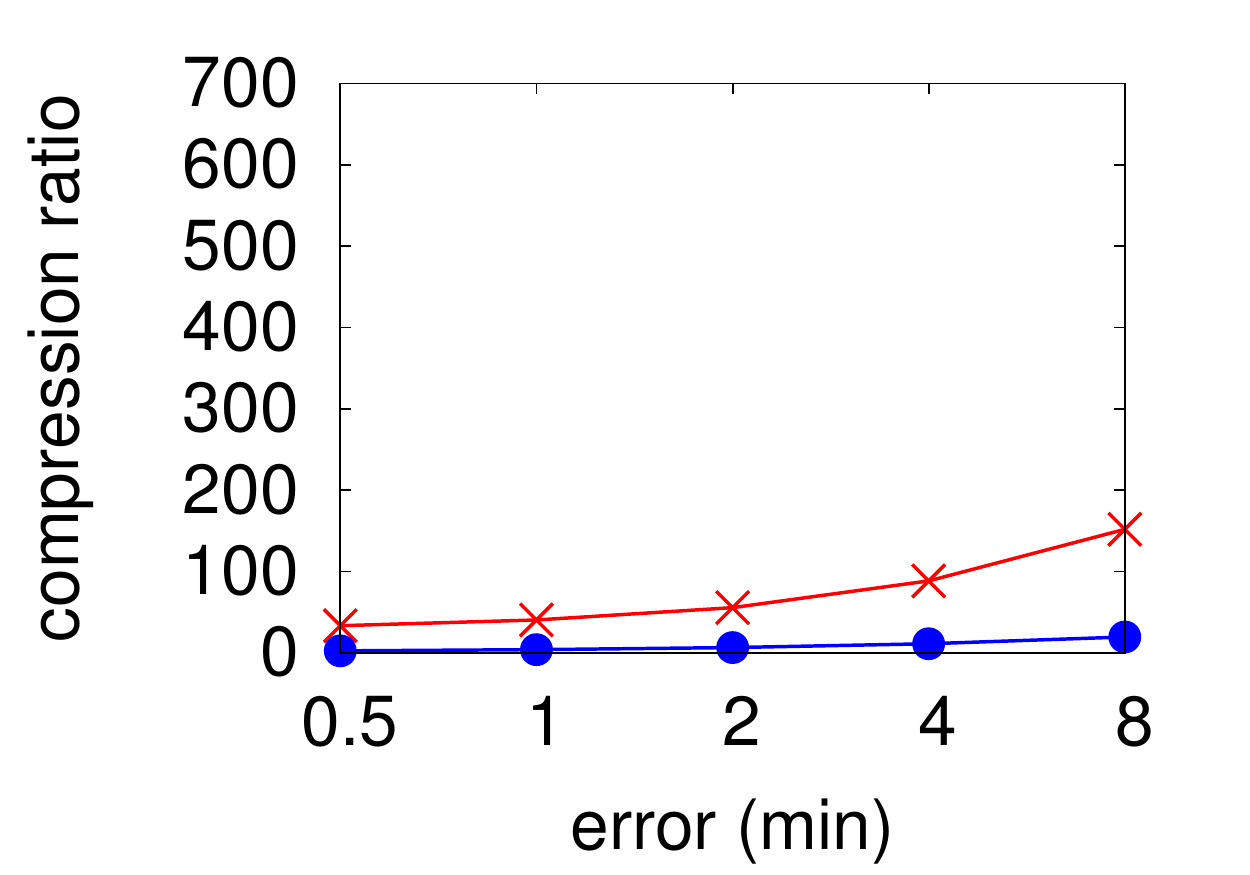}
}
\subfloat[SF \label{fig::temporal_comp_time_sf}]{
\includegraphics[width=0.2\textwidth]{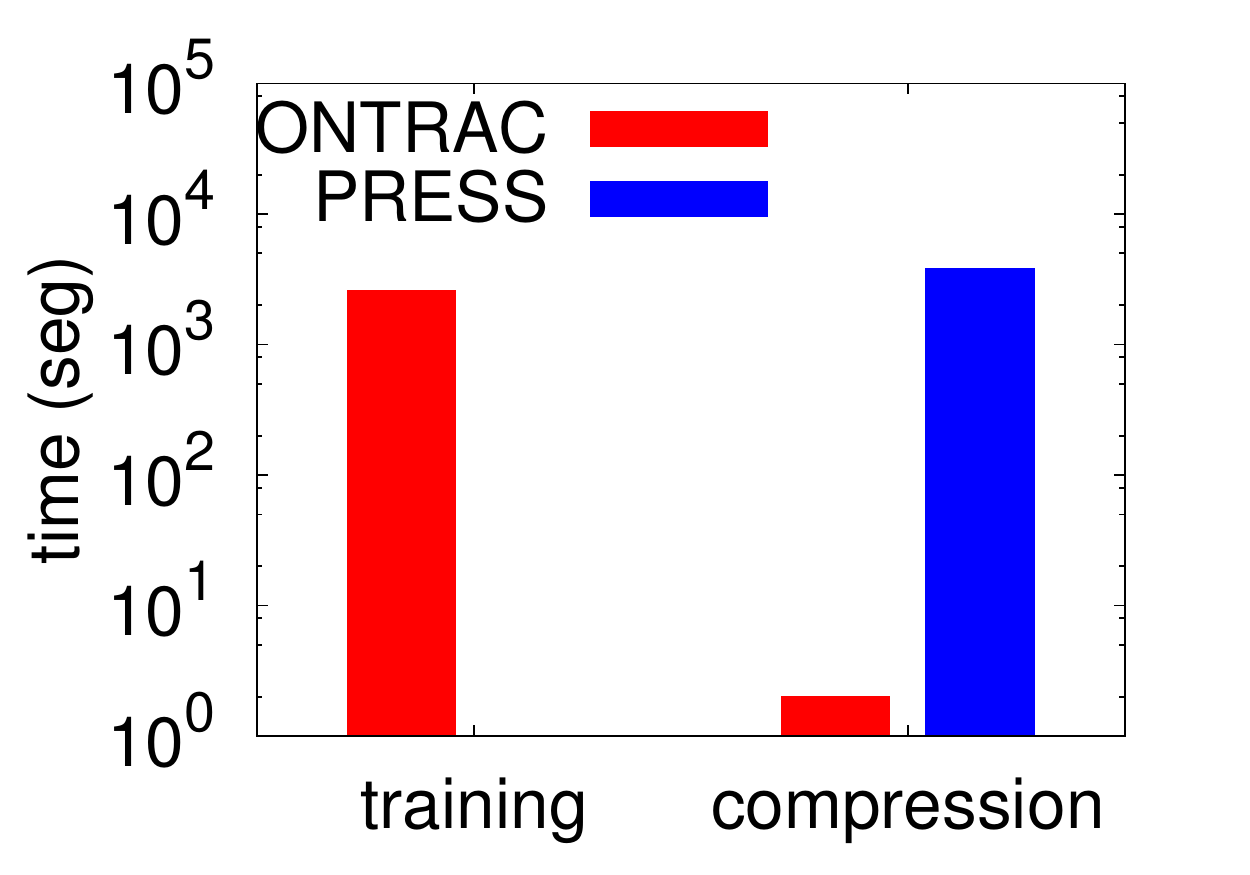}
}
\subfloat[Beijing \label{fig::temporal_comp_time_beijing}]{
\includegraphics[width=0.2\textwidth]{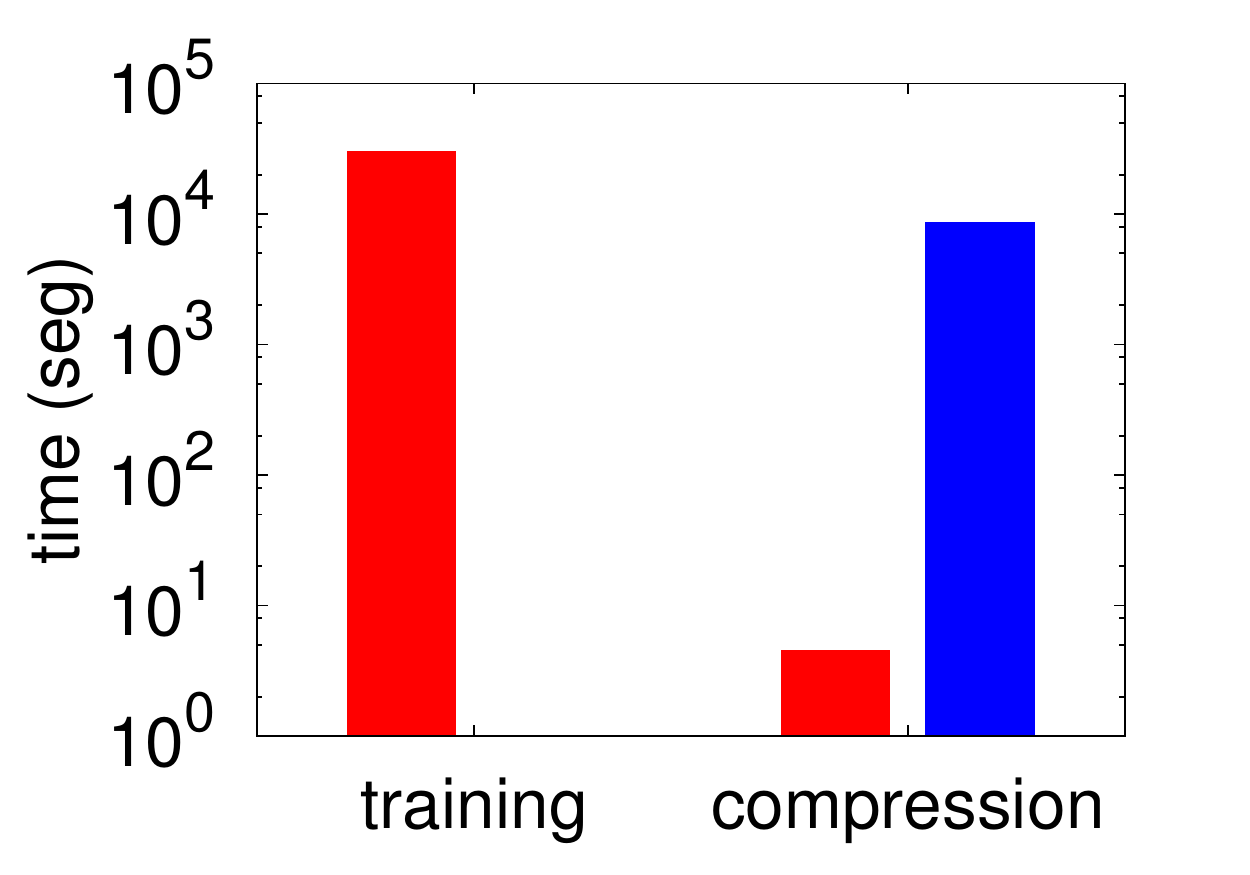}
}
\caption{Temporal compression ratio (a,b), and training and compression times (c,d) for ONTRAC and PRESS using the SF (a, c) and Beijing (b, d)  datasets. ONTRAC achieves up to 3 times higher compression ratio than PRESS. Our approach is efficient in terms of compression time, significantly outperforming the baseline. Efficient compression is of particular interest in the online trajectory compression setting. \label{fig::temporal_compression_results}}
\end{figure*}

\begin{figure*}[ht!]
\centering
\subfloat[SF \label{fig::inserts_sf}]{
\includegraphics[width=0.2\textwidth]{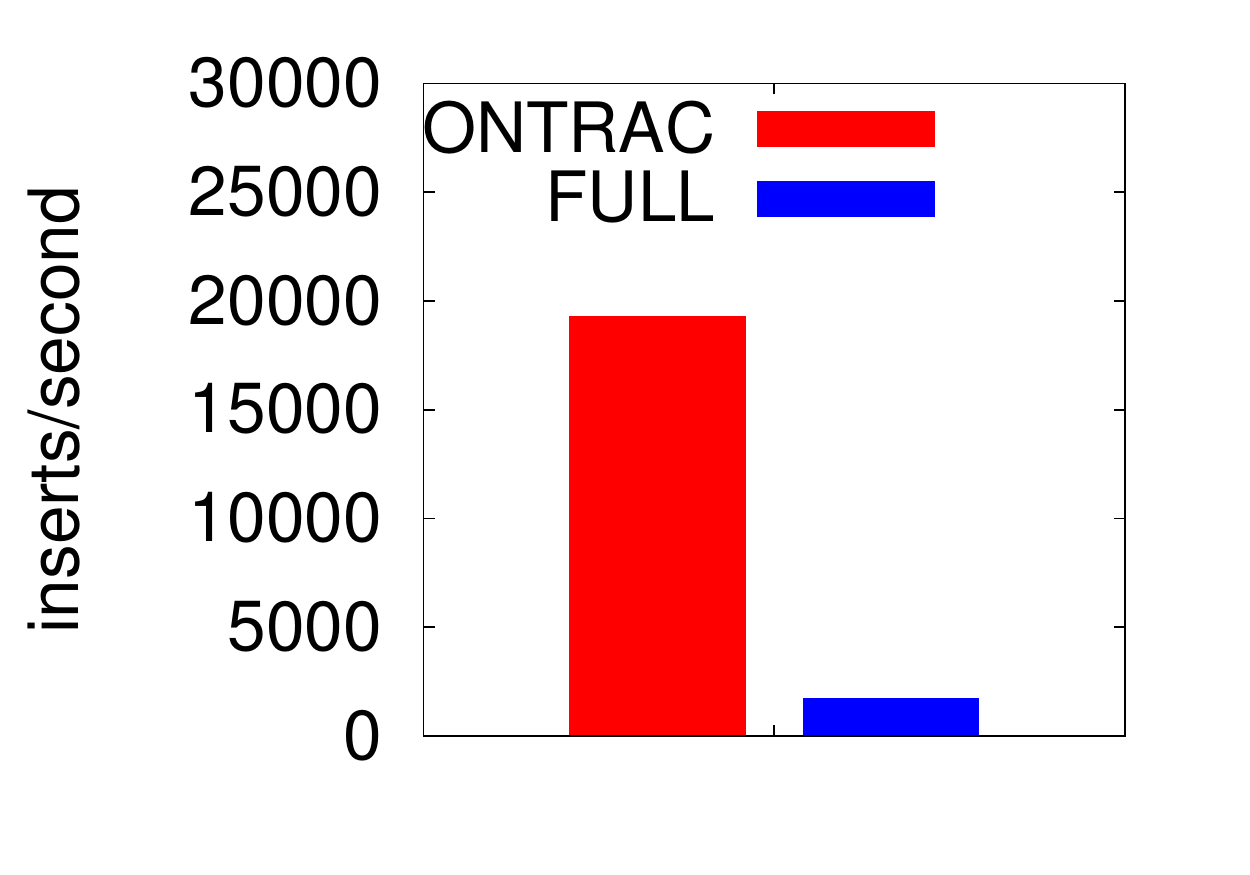}
}
\subfloat[Beijing \label{fig::inserts_beijing}]{
\includegraphics[width=0.2\textwidth]{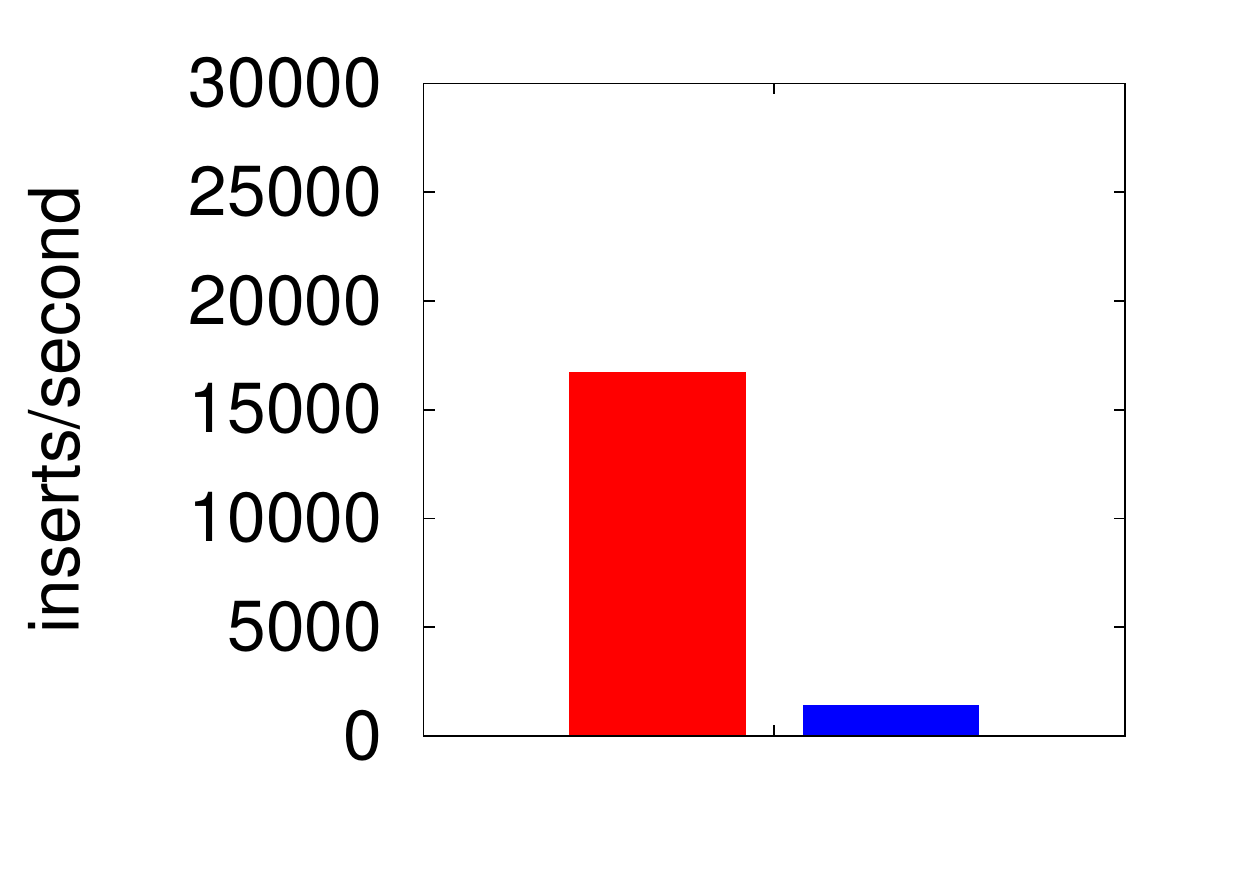}
}
\subfloat[SF \label{fig::query_sf}]{
\includegraphics[width=0.2\textwidth]{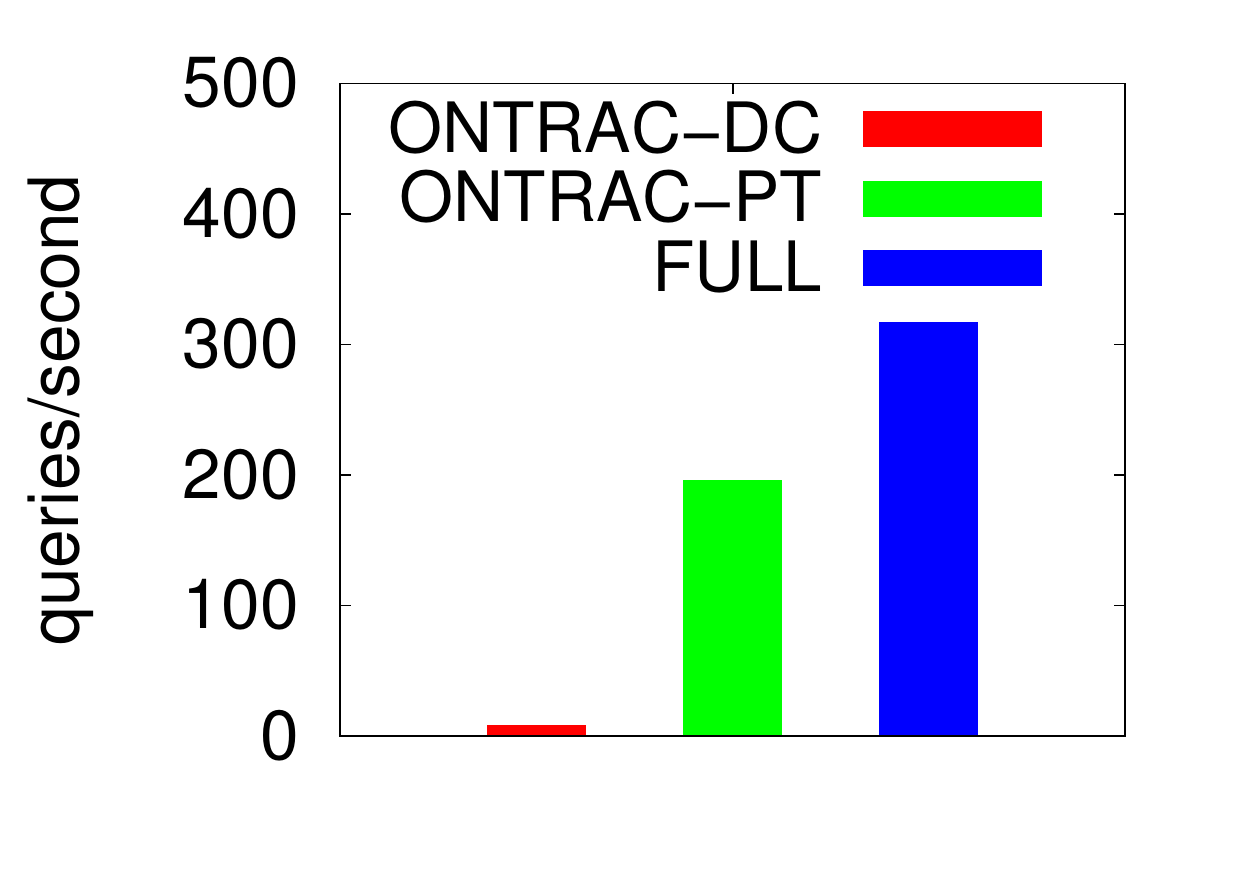}
}
\subfloat[Beijing \label{fig::query_beijing}]{
\includegraphics[width=0.2\textwidth]{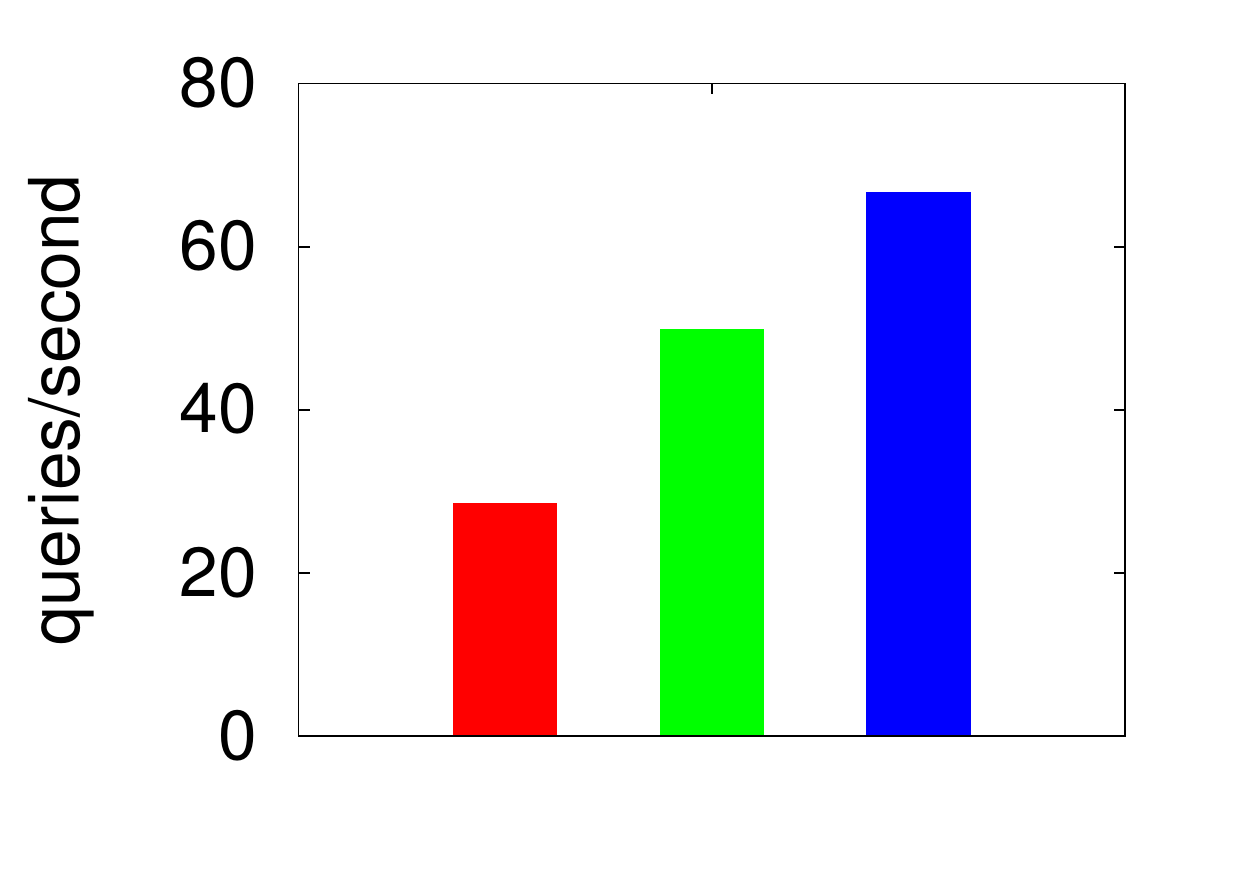}
}
\caption{Inserts (a,b) and queries/second (c,d) when applying ONTRAC with complete (ONTRAC-DC) and partial decompression (ONTRAC-PT) and with no compression (FULL) using SF (a,c) and Beijing (b,d) datasets. ONTRAC increases database scalability by up to one order of magnitude with a small overhead on query performance. Also, ONTRAC-PT achieves significantly lower query times compared to ONTRAC-DC for long trajectories (SF and Beijing).}
\end{figure*}

\subsection{Temporal Compression}
\label{subsec::exp_temporal_compression}

In this section, we evaluate ONTRAC and the baseline technique in terms of temporal compression. Our main goal is to assess how many updates to the database can be suppressed by applying ONTRAC's prediction scheme. As in the last section, we divide the datasets into training and test folds with 90\% and 10\% of the trajectories, respectively. Moreover, we vary the accepted error $\lambda$ from .5 to 8 minutes as means to analyze the trade-off between compression accuracy and compression ratio in the lossy compression. 

The relationship between the compression error ($\lambda$) and the compression ratio for ONTRAC and PRESS is shown in Figures \ref{fig::temporal_comp_ratio_sf} and \ref{fig::temporal_comp_ratio_beijing}. ONTRAC outperforms PRESS for all the datasets and different values of $\lambda$. For instance, ONTRAC achieves from $16$ ($\lambda=0.5$) to $21$ ($\lambda=8$) times higher compression for \textit{SF} dataset and from $8$ ($\lambda=8$) to $14$ ($\lambda=0.5$) higher compression for \textit{Beijing}.   

In Figures \ref{fig::temporal_comp_time_sf} and \ref{fig::temporal_comp_time_beijing}, we evaluate ONTRAC and PRESS compression with regard to training and compression time. PRESS performs line simplification, which requires no training. On the other hand, ONTRAC's \textbf{Temporal-training} is an EM algorithm and thus performs several iterations to compute its prediction model. Results show that our training algorithm requires approximately 3 hours for training using 12 million trajectories and 8 cores. Because ONTRAC's \textit{e-step} can be performed in parallel for multiple trajectories, it can easily exploit a larger number of cores to achieve lower training times. Opposite results are seen for compression time, where ONTRAC outperforms PRESS by up to two orders of magnitude. Efficient compression is of particular interest in the online compression setting.

\subsection{Scalability and Querying}
\label{subsec::exp_querying_storage}

Lastly, we evaluate how ONTRAC affects database scalability and querying performance. The scenario considered is similar to the smart city application used as a motivating example for online trajectory compression (see Figure \ref{fig::ontrac}), where updates are inserted into a database in an online manner and queries are performed in order to retrieve some information from these trajectories. ONTRAC's compression models are learned using 90\% of the trajectories for training. Scalability is measured in terms of the number of \textit{inserts} processed per second and querying performance is evaluated in terms of the number of \textit{where} (Definition \ref{def::where_query}) queries processed per second. We compare our compression approach against the no-compression setting.

Figures \ref{fig::inserts_sf} and \ref{fig::inserts_beijing} show the database scalability using ONTRAC ($\lambda=1$ min.) and without any compression (FULL). ONTRAC increases the number of \textit{inserts}/second by factors of $11.0$ and $11.7$ for \textit{SF} and \textit{Beijing} datasets, respectively. We also evaluate the impact of ONTRAC over querying performance, for complete (ONTRAC-DC) and partial (ONTRAC-PT) trajectory decompression, in Figures \ref{fig::query_sf} and \ref{fig::query_beijing}. ONTRAC-DC achieves poor query performance due to the cost of full decompression. This was the main motivation for the partial decompression approach (ONTRAC-PT), which incurs limited overheads of $38$\% and $25$\% compared to the FULL setting for \textit{SF} and \textit{Beijing}, respectively.

%% file: 2_related_work.tex
\section{Related Work}

Our work is focused on map-matched trajectory compression, a problem introduced in \cite{Cao:2005:NMI:2131560.2131576}. The proposed approach, called Nonmaterialized Trajectory Construction (NTC), represents trajectories in a linear reference system (distance, time) and finds a minimum sequence of line segments that covers the original trajectory within a user-defined error. Map-matched trajectory compression (MMTC) \cite{kellaris2013map} compresses a trajectory by replacing some of its parts by shorter sub-trajectories (i.e. those with fewer road segments). 

In a recent work \cite{pvldbSongSZZ14}, the authors proposed proposed decomposing trajectories into spatial and temporal components for which different compression schemes were presented. The resulting approach, called PRESS, applies shortest paths and Huffman coding to compress the spatial information and function simplification (similar to \cite{Cao:2005:NMI:2131560.2131576}) for the temporal information. Experiments show that it outperforms both NTC and MMTC in terms of compression ratio and query performance. However, PRESS relies on a batch model, an aspect that limits its application in online settings, such as taxi dispatching \cite{liao2003real} and traffic monitoring \cite{Guo:2014:TTT:2694428.2694432}, where large volumes of trajectories have to be processed in real-time in order to support high-quality services. Moreover, PRESS compressed data cannot be directly processed by existing relational databases, which offer many capabilities (e.g. indexing) that are critical for trajectory data management. These limitations are the motivation for our work. 


Similar to PRESS, ONTRAC also employs different compression schemes for spatial and temporal information. More specifically, its spatial compression is inspired by \textit{Prediction by Partial Matching (PPM)} \cite{1096090}, a state-of-art statistical compression scheme. PPM enables the creation of compact bit string representations for symbols that are more likely to occur given a sequence of previous symbols. In ONTRAC, we apply a Markov model to suppress the most likely update for an object from the database based on the road segments it has traversed recently. Similar models have been recently applied for the inference of uncertain trajectories \cite{banerjee2014inferring} and for answering predictive queries on road networks \cite{hendawi2015predtree}.

ONTRAC's temporal compression exploits \textit{Quadratic}  {Programming (QP)} \cite{boyd2004convex} and \textit{Expectation Maximization (EM)} \cite{dempster1977maximum} in order to learn a travel-time model to predict future updates. While there is an extensive literature on travel-time inference \cite{hunter2009path,ide2011trajectory,kim2007path,yang2013travel,wang2014travel,yuan2011driving}, the online compression setting has important properties that motivate a new approach. For instance, the inference has to be made at update time, which imposes higher performance constraints. Moreover, online compression requires shorter term predictions since travel-times will be in the order of minutes. 


%% file: 6_conclusion.tex
\section{Conclusion}

We have described ONTRAC, a new framework for online map-matched trajectory compression. ONTRAC applies prediction models for spatial and temporal trajectory information as means to suppress updates to a trajectory database. In particular, ONTRAC's spatial compression predicts road segments  via a Markov model and its temporal compression scheme learns a travel-time model by combining Quadratic Programming and Expectation Maximization. 

Experiments have shown that our approach achieves high compression ratios in terms of spatial and temporal compression for real datasets. ONTRAC outperforms the baseline method even when 4 minute delays in updates are allowed and achieves up to 21 times higher compression for travel-times. Finally, we have shown that ONTRAC can increase database scalability by up to one order of magnitude with small overhead over querying performance.

This paper opens promising directions for future research. Adding support to other, more complex queries, such as nearest-neighbors and range queries, using partial reconstruction would increase the applicability of our framework. Also, a context-based travel-time prediction, similar to our spatial compression scheme, might improve ONTRAC's compression power. Finally, online EM \cite{neal1998view} can support efficient online updates of ONTRAC's temporal compression model.

%% file: appendix.tex
\section{Proofs of Theorems}

\textbf{Theorem \ref{thm:network}}
\begin{proof}
Given the random-walk trajectory assumption:
\begin{align}
h_k &= 1 - \sum_{S' \in \mathcal{D}} P(S'\sqcup \Psi(S',\mathcal{D})) \nonumber \\
    &= 1 - \sum_{S' \in \mathcal{D}} P(s_k \sqcup \Psi(S',\mathcal{D})) \nonumber \\
    &= h_1 \nonumber 
\end{align}
We know that $P(s_i)$ is the PageRank $\pi(s_i)$ of $s_i$ in $G$. Finally, again using the random-walk assumption:
$$P(s_i\sqcup s_j) = \pi(s_i)\frac{1}{deg^o(s_i)}, \forall (s_i,s_j) \in E$$
\end{proof}

\textbf{Theorem \ref{thm:qp_form}}
\begin{proof}
From maximum-likelihood theory, we know that:
$$T_{max}'=\max_{T'}P(T')=\min_{T'}\{-\log P(T')\}$$

Also, from the Gaussian distribution definition:

$$N(x;\mu,\sigma)=\frac{1}{\sigma \sqrt{2\pi}} e^{-\frac{(x-\mu)^2}{2\sigma^2}}$$
$$\log N(x;\mu,\sigma)=-\log \sigma -\frac{(x-\mu)^2}{2\sigma^2} -\frac{1}{2} \log 2\pi$$

After removing some constants, we can write the logarithms of the probabilities in Equation \ref{eqn:traj_like} as follows:

\begin{scriptsize}
\begin{align}
\log P(t_i'|\phi_i,\omega_i) &= -\frac{(t_i'-\phi_i)^2}{2\omega_i^2} \label{eqn:log_model}\\
\log P(t_i'|t_{i-1}',\Delta) &= -\frac{1}{2\Delta^2} (\frac{t_i'}{|s_i|}-\frac{t_{i-1}'}{|s_{i-1}|})^2 \label{eqn:log_smooth}\\
\log P(\sum_{i=(j,1)}^{(j,k)} t_i' | \overline{t}_j,\sigma^{\star}) &= -\frac{(\sum_{i=(j,1)}^{(j,k)}t_i'-\overline{t}_j)^2}{2\sigma_j^2} \label{eqn:log_gps}
\end{align}
\end{scriptsize}

And $\log P(T')$ can be computed as:
\begin{scriptsize}
$$\sum_{i = (1,1)}^{(n,k)} \log P(t_i'|\phi_i,\omega_i) + \sum_{i={(1,2)}}^{(n,k)} \log P(t_i'|t_{i-1}') + \sum_{j=1}^{n} \log P(\sum_{i=(j,1)}^{(j,k)} t_i' | \overline{t}_j)$$
\end{scriptsize}
where some parameters are omitted for conciseness. Expanding Equation \ref{eqn:log_model} and dropping some constants:

\begin{scriptsize}
$$\sum_{i = (1,1)}^{(n,k)} \log P(t_i'|\phi_i,\omega_i) = -\sum_{i = (1,1)}^{(n,k)} \frac{t_i'^2}{2\omega_i^2} + \sum_{i = (0,0)}^{(n,k)} \frac{t_i'\phi_i}{\omega_i^2}$$
\end{scriptsize}

We represent times in vector form $\textbf{x}$ ($\textbf{x}[i]=t_i'$) and take the $\frac{1}{2}$ factor in the QP notation into account, which leads to $Q_1$ and  $\textbf{c}_1$. Doing the same for Equations \ref{eqn:log_smooth} and \ref{eqn:log_gps}:

\begin{scriptsize}
\begin{align}
\sum_{i={(1,1)}}^{(n,k)} \log P(t_i'|t_{i-1}',\Delta) &= -\frac{t_{1,1}'^2}{2\Delta^2|s_{1,1}|^2} -\frac{t_{n,k}'^2}{2\Delta^2|s_{n,k}|^2} \nonumber\\
							& - \sum_{i=(1,2)}^{(n,k)-1} \frac{t_i'^2}{\Delta^2|s_i|^2} + \sum_{i=(1,2)}^{(n,k)} \frac{t_i't_{i-1}'}{\Delta^2|s_i||s_{i-1}|}\nonumber
\end{align}
\end{scriptsize}
\begin{scriptsize}
\begin{align}
\sum_{j=1}^{n} \log P(\sum_{i=(j,1)}^{(j,k)} t_i' | \overline{t}_j,\sigma^{\star}) &= -\sum_{j=1}^{n} \frac{\sum_{i=(j,1)}^{(j,k)} t_i'^2}{2\sigma_j^2} \nonumber\\
 &-\sum_{j=1}^{n} \frac{\sum_{i=(j,1)}^{(j,k)} \sum_{\ell=i+1}^{(j,k)}t_i't_{\ell}'}{\sigma_j^2} \nonumber\\
 &+\sum_{j=1}^{n} \frac{\overline{t}_j\sum_{i=(j,1)}^{(j,k)} t_i'}{2\sigma_j^2} \nonumber
\end{align}
\end{scriptsize}

These are the expressions for computing $Q_2$, $Q_3$ and $\textbf{c}_2$.
\end{proof}

\textbf{Theorem \ref{thm:convex}}
\begin{proof}
A sufficient condition for the convexity of the formulation is $Q$ to be positive-definite (i.e. $\textbf{x}Q\textbf{x}^T > 0$ for any non-zero vector $\textbf{x}$) \cite{boyd2004convex}. From the definition of $Q$:
\begin{align}
\textbf{x}Q\textbf{x}^T&= \textbf{x}(Q_1+Q_2+Q_3)\textbf{x}^T\nonumber\\
 & = \textbf{x}Q_1\textbf{x}^T+\textbf{x}Q_2\textbf{x}^T+\textbf{x}Q_3\textbf{x}^T\nonumber
\end{align}
$Q_1$ is positive-definite because it is a diagonal matrix with positive entries, its eigenvalues are the diagonal. $Q_2$ is positive-definite due to the fact that it is a block diagonal matrix in which each $2\times2$ block is itself a positive-definite matrix. $Q_3$ is also a block diagonal matrix where each block is constant and thus positive-semidefinite, the same holding for $Q_3$  (i.e. $\textbf{x}Q_3\textbf{x}^T \geq 0$, for any non-zero $\textbf{x}$).
\end{proof}

\textbf{Theorem \ref{thm::convergence}}
\begin{proof}
We will prove the convergence of the algorithm based on how new estimate $(\Phi^{\ell},\Omega^{\ell})$ is related to the one computed in the previous iteration, $(\Phi^{\ell-1},\Omega^{\ell-1})$:
\begin{equation}
(\Phi^{\ell},\Omega^{\ell}) = \argmax_{\Phi,\Omega} \mathcal{Q}(\Phi,\Omega,\Phi^{\ell-1},\Omega^{\ell-1})
\end{equation}
where (replacing $t_i'$ by $t_i$ for conciseness):
\begin{small}
\begin{equation}
\mathcal{Q}  = \sum_{T \in \mathcal{D}_t} \sum_{i=1}^{|T|} \int_{t_i} P(t_i| \Phi^{\ell-1}, \Omega^{\ell-1})\log P(t_i|\Phi,\Omega)dt_i \nonumber
\end{equation}
\end{small}
We maximize $\mathcal{Q}$ by computing $\Phi$ and $\Omega$ such that $\frac{d\mathcal{Q}}{d \Phi}= 0$ and $\frac{d\mathcal{Q}}{d \Omega} = 0$ (dropping some constants): 
\begin{footnotesize}
\begin{align}
\frac{d\mathcal{Q}}{d \Phi} &= -\sum_{T \in \mathcal{D}_t}^n \sum_{i=1}^{|T|} \int_{t_i} P(t_i|\Phi^{\ell}, \Omega^{\ell}) \frac{d}{d\phi_i}(t_i-\phi_i)^2dt_i \nonumber\\
&= -\frac{1}{|\mathcal{T}_i|}\sum_{s_i \in \mathcal{T}}^{|\mathcal{T}|}t_i + \phi_i\nonumber
\end{align}
\end{footnotesize}
Where $\mathcal{T}_i = \{T \in \mathcal{D}_t | s_i \in T\}$. Similarly:
\begin{scriptsize}
\begin{align}
\frac{d\mathcal{Q}}{d \Omega} &= \sum_{T \in \mathcal{D}_t}^n \sum_{i=1}^{|T_i|} \int_{t_i} P(t_i|\Phi^{\ell}, \Omega^{\ell}) \frac{d}{d\omega} \left(-\frac{(t_i-\phi_i)^2}{2\omega_i^2}-\log \omega_i\right) dt_i\nonumber
\end{align}
\end{scriptsize}
By solving the derivative and eliminating some multiplicative terms, we get an equivalent expression:
\begin{small}
\begin{equation}
\frac{d\mathcal{Q}'}{d\Omega} = \sum_{j=1}^{|\mathcal{T}_i|} (t_i-\phi_i)^2 - \mathcal{T}\omega^2\nonumber
\end{equation}
\end{small}
This proves that lines 10 and 11 of \textbf{Temporal-training} cannot decrease the likelihood of the trajectories and thus the algorithm is guaranteed to converge.
\end{proof}

\section{Additional Experiments for Real and Synthetic Data}

\begin{table}[t!]
\centering
\scriptsize
\begin{tabular}{| c | c | c | c | c |}
\hline
\textbf{Syn-I} & 78,847 & 181,598 & 3.3M& 2m\\
\hline
\textbf{Syn-II} & 78,847 & 181,598 & 0.7M& 10m\\
\hline
\end{tabular}
\caption{Synthetic dataset statistics. \label{table:syn_data_stats}}
\end{table}

\subsection{Synthetic Data}

Experiments were also performed using two synthetic datasets (\textbf{Syn-I} and \textbf{Syn-II}) generated using the \textit{SF} road network and synthetic trajectories. Each trajectory was created by selecting a random starting and destination segments. Road segments in a trajectory are those in the shortest path between the starting point and destination. Travel-times follow segment distributions as defined in our trajectory model (see Section \ref{sec:gaussian_model}) with Gaussian distributed speeds from $N(15,10)$ (in m/s). Destinations are selected  according to an exponential popularity distribution ($\alpha e^{-\alpha x}$), with $\alpha=1$ in \textbf{Syn-I} and $\alpha=10^{-4}$ in \textbf{Syn-II}. Therefore, while \textbf{Syn-I} has a few very popular destinations, \textbf{Syn-II} has many almost equally popular ones.

\subsection{Network and Trajectory Entropies}

\begin{figure}[t!]
\subfloat[Network \label{fig::net_entropy}]{
\includegraphics[width=0.2\textwidth]{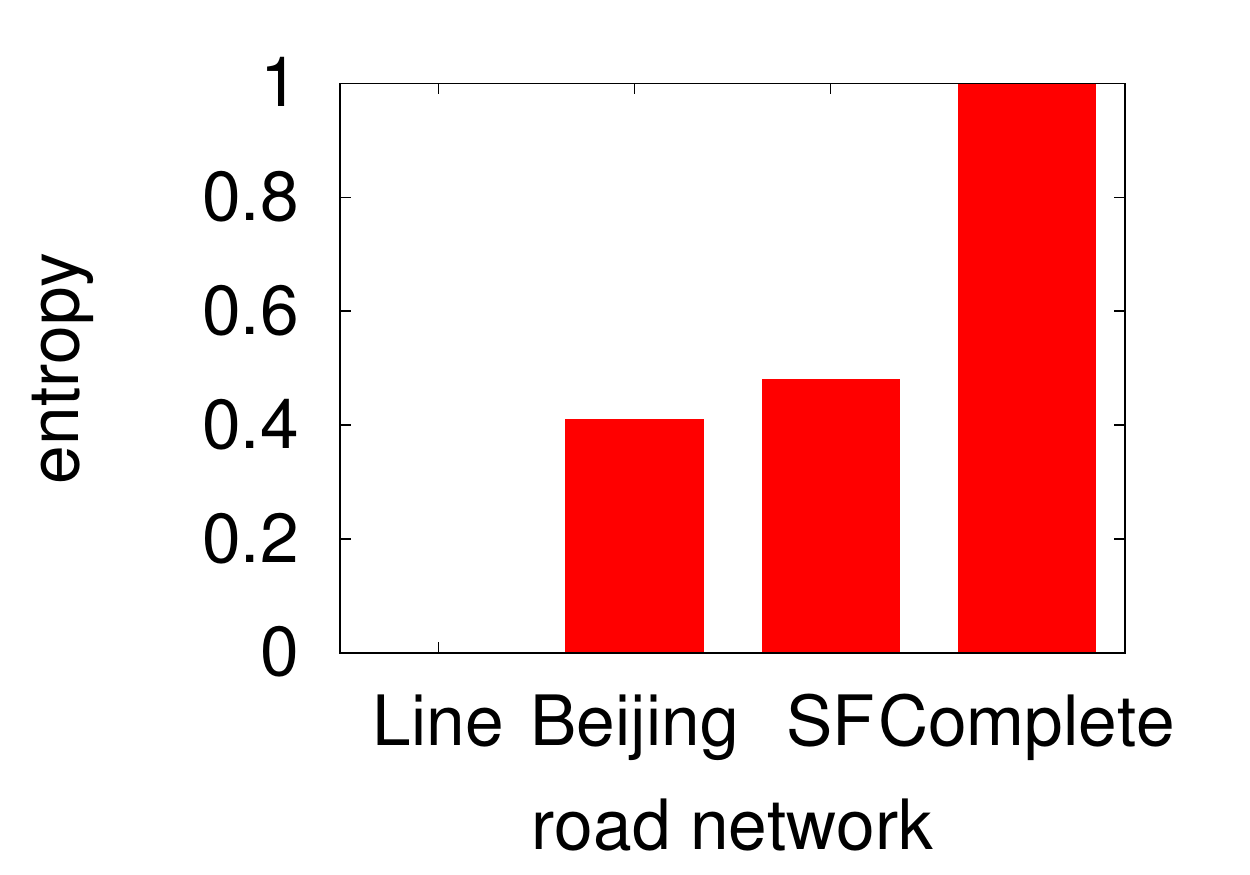}
}
\subfloat[Trajectory \label{fig::traj_entropy}]{
\includegraphics[width=0.2\textwidth]{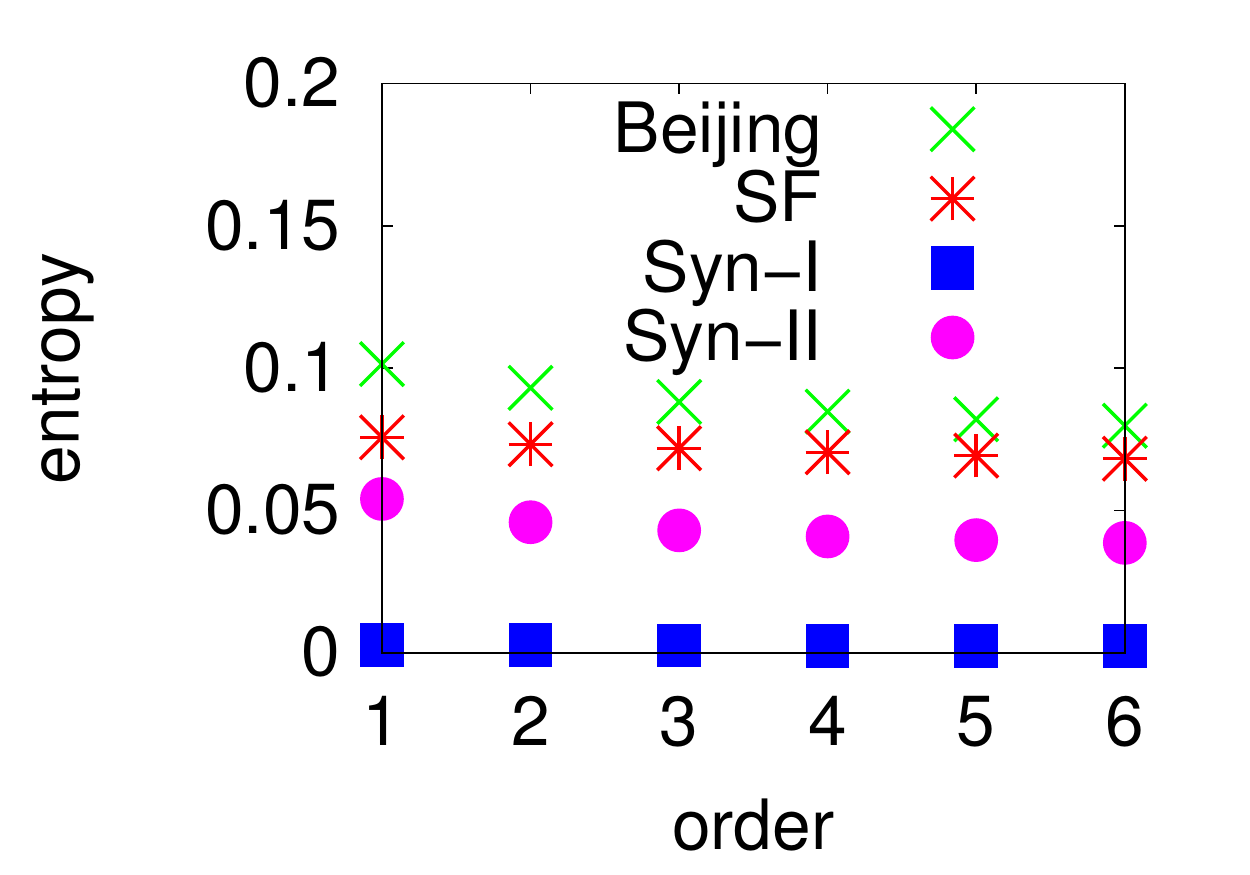}
}
\caption{Network and trajectory entropies.}
\end{figure}

We characterize the \textit{SF} and \textit{Beijing} datasets in terms of \textit{spatial update block entropy}, as defined in Section \ref{subsec::spatial_compression}. Figure \ref{fig::net_entropy} shows the entropy of the networks (see Theorem \ref{thm:network}) for random-walk trajectories and compares them against a line and a complete graph. Both road networks are sparse and the results show that more than half of the trajectory updates can be saved in the extreme case of random trajectories. \textit{SF} has an entropy 17\% larger than \textit{Beijing}.

Figure \ref{fig::traj_entropy} compares the trajectories from datasets in terms of entropy for varying order $k$. As expected, the entropy decreases as $k$ grows, but differences between entropies at order 1 and order 6 are small (within 0.03). Trajectories in the \textit{Beijing} dataset present larger entropy than those in \textit{SF}. Also, by design, trajectories in \textit{Syn-II} have higher entropy than those in \textit{Syn-I}. Based on these results, we fix  ONTRAC's spatial compression order to 2 in the remaining of the experiments discussed in this section. 

\subsection{Spatial Compression}

For \textit{Syn-I}, where most of the trajectories share the same destination, our approach achieves a compression ratio of 54. However, in \textit{Syn-II}, where destination popularity is almost uniformly distributed, its compression ratio decreases to 16.

\begin{figure*}[ht!]
\centering
\subfloat[Syn-I \label{fig::spatial_comp_ratio_syn_one}]{
\includegraphics[width=0.2\textwidth]{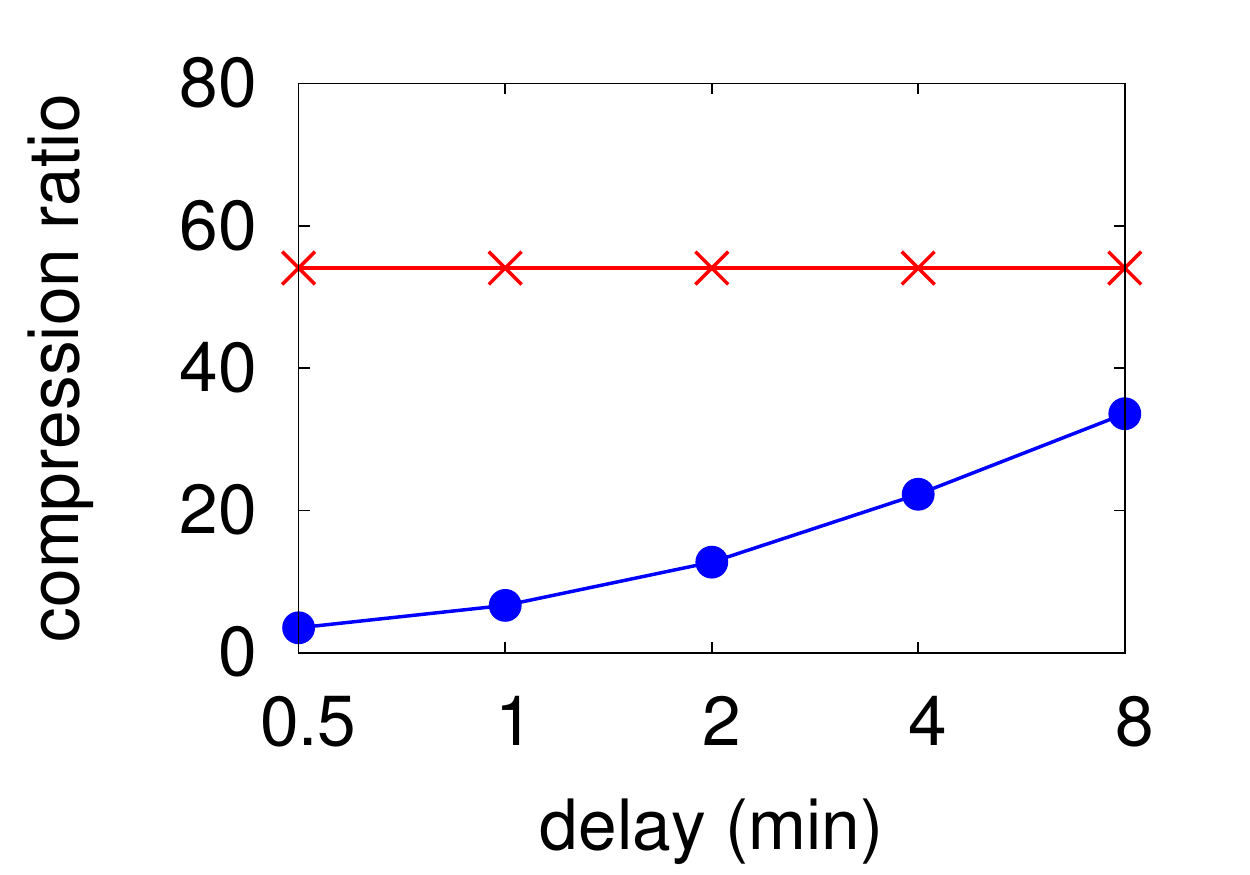}
}
\subfloat[Syn-II \label{fig::spatial_comp_ratio_syn_two}]{
\includegraphics[width=0.2\textwidth]{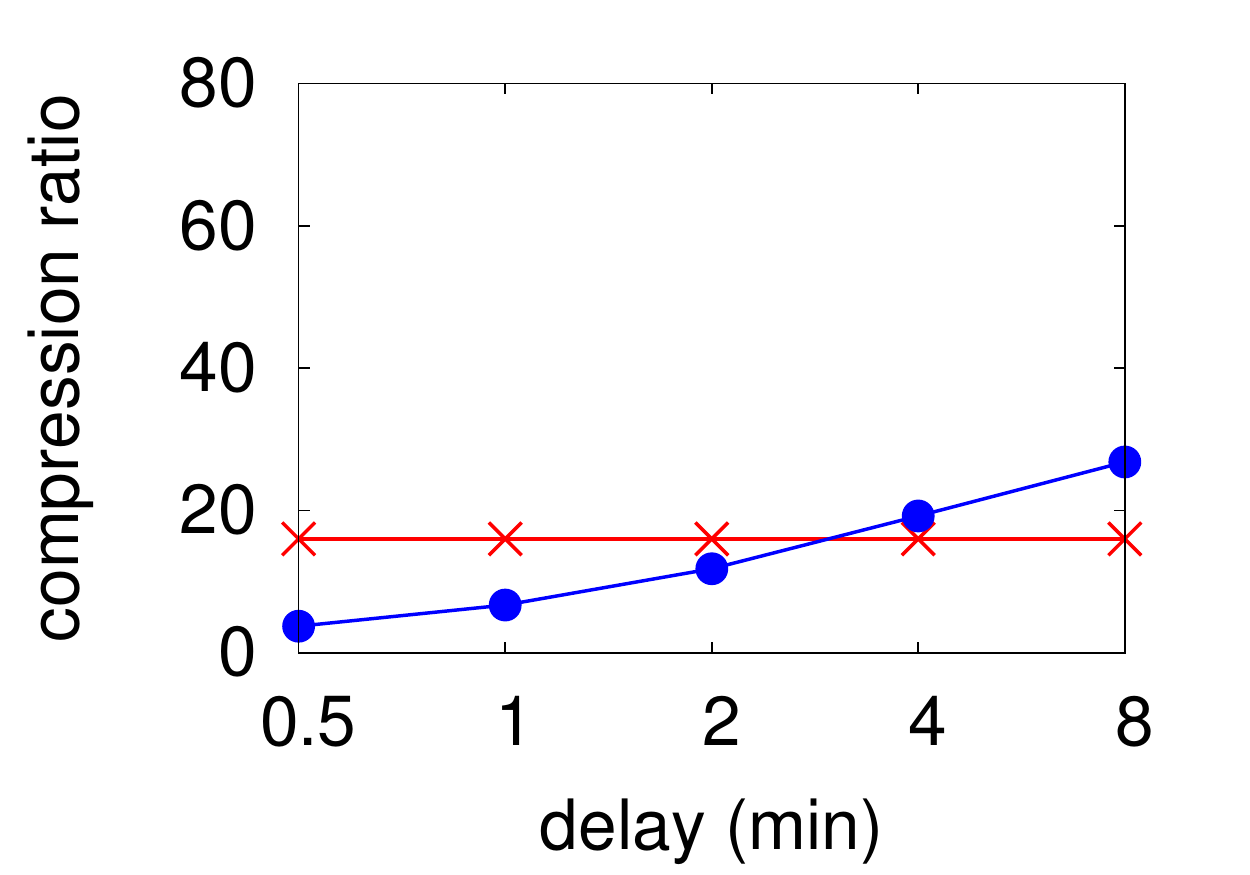}
}
\subfloat[Syn-I \label{fig::spatial_comp_time_syn_one}]{
\includegraphics[width=0.2\textwidth]{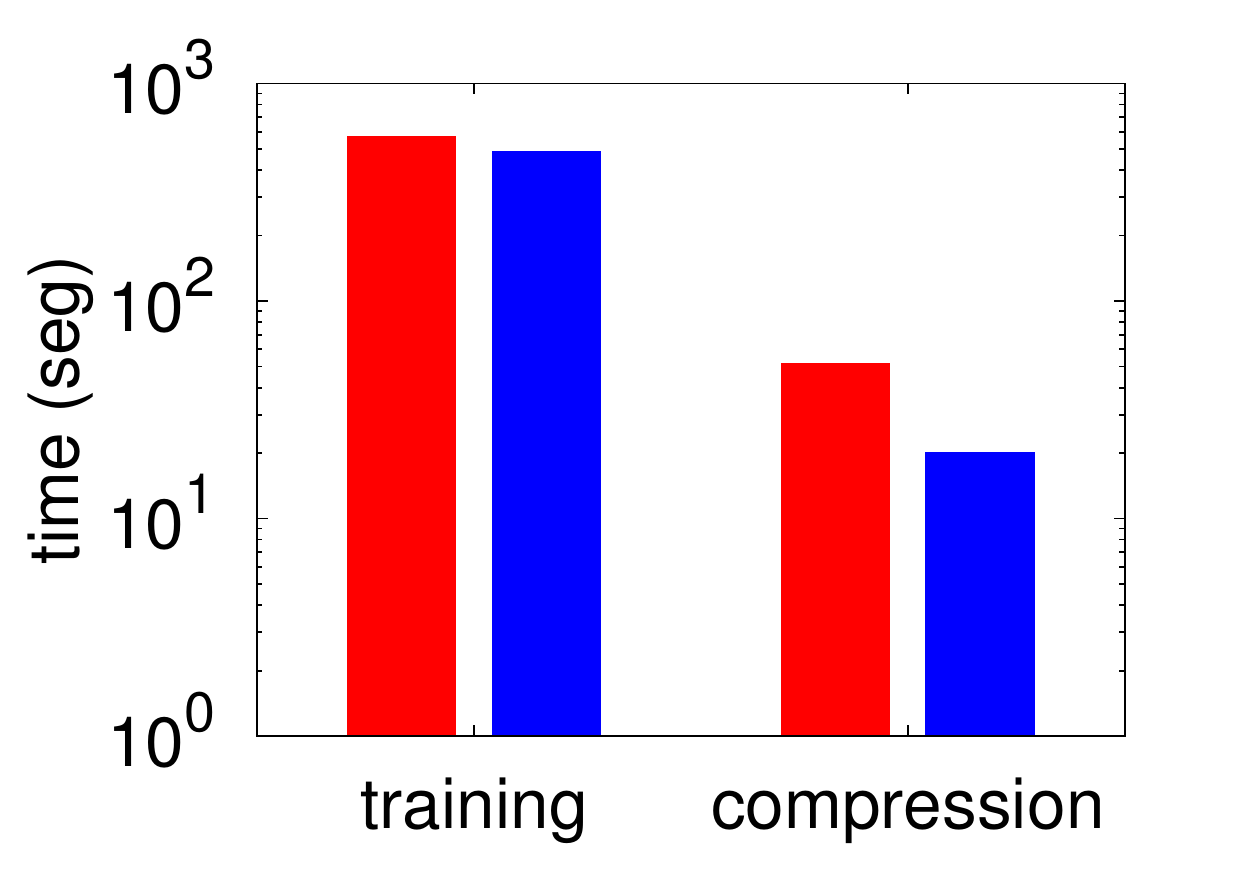}
}
\subfloat[Syn-II \label{fig::spatial_comp_time_syn_two}]{
\includegraphics[width=0.2\textwidth]{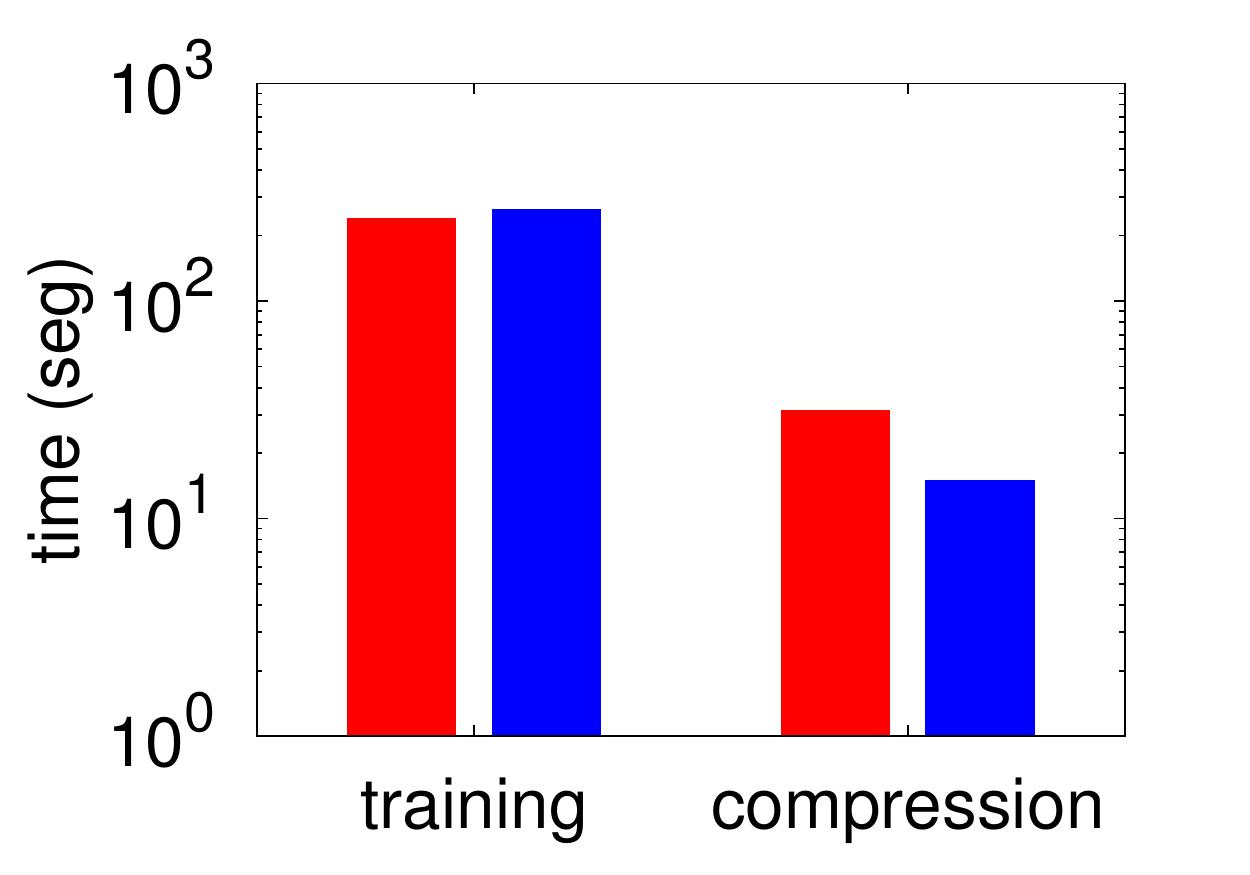}
}
\caption{Spatial compression ratio (a,b), training and compression times (c,d) for ONTRAC and PRESS using Syn-I (a,c), and Syn-II (b,d) datasets. \label{fig::syn_spatial_compression_results}}
\end{figure*}

\subsection{Temporal Compression}

\begin{figure*}[ht!]
\centering
\subfloat[SF \label{fig::temporal_conv_sf}]{
\includegraphics[width=0.2\textwidth]{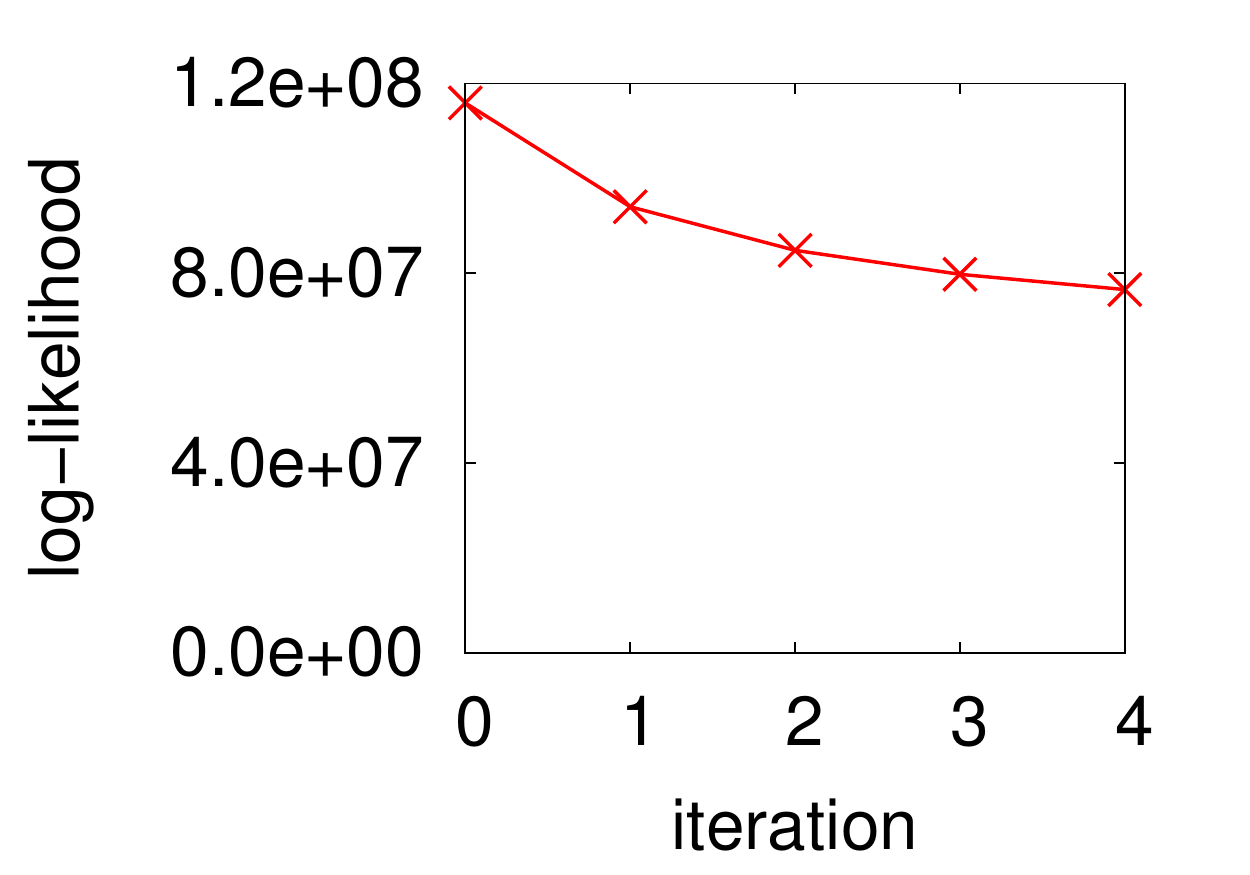}
}
\subfloat[Beijing \label{fig::temporal_conv_beijing}]{
\includegraphics[width=0.2\textwidth]{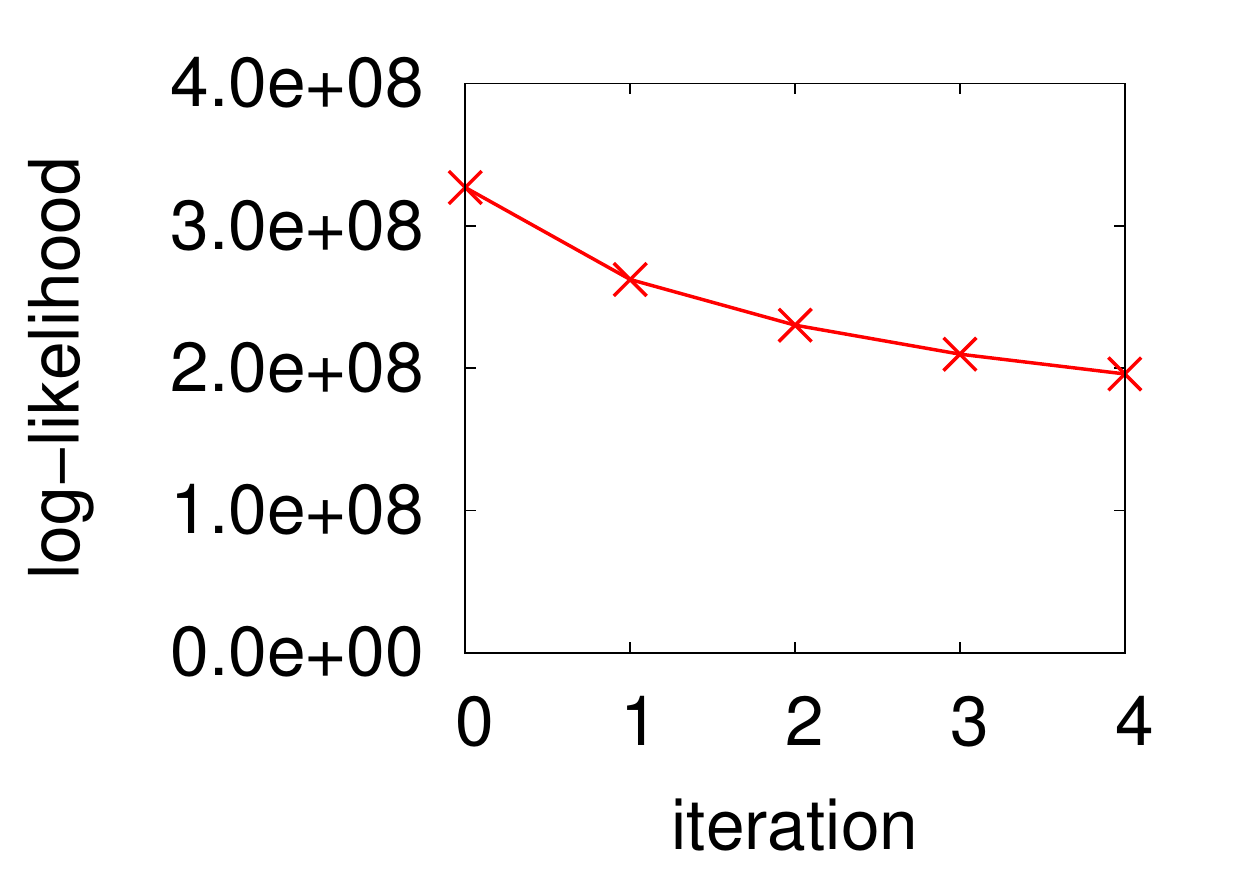}
}
\subfloat[Syn-I \label{fig::temporal_conv_syn_one}]{
\includegraphics[width=0.2\textwidth]{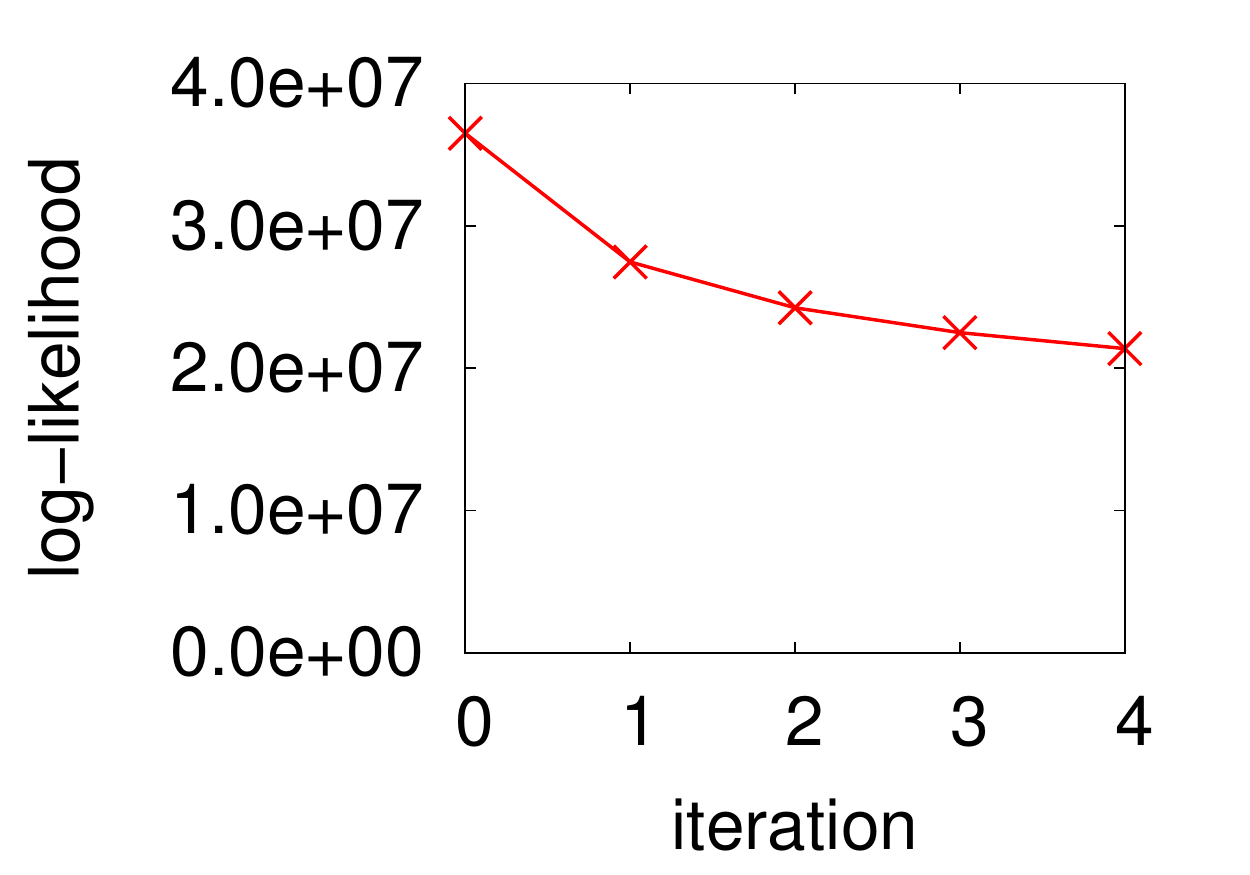}
}
\subfloat[Syn-II \label{fig::temporal_conv_syn_two}]{
\includegraphics[width=0.2\textwidth]{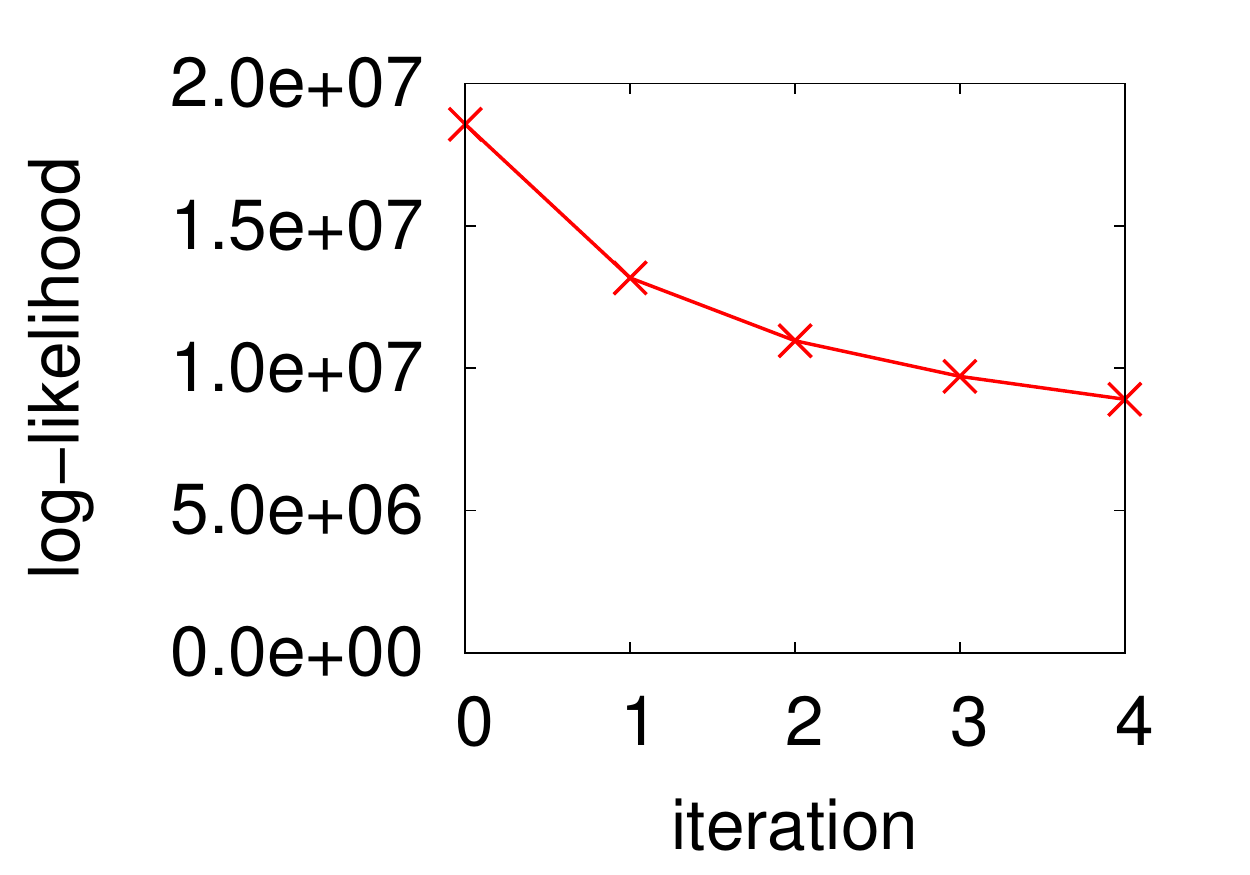}
}
\caption{Temporal compression convergence using the SF (a), Beijing (b), Syn-I (c), and Syn-II (d) datasets.}
\end{figure*}

Figures \ref{fig::temporal_conv_sf}-\ref{fig::temporal_conv_syn_two} provide experimental evidence of the convergence of our approach (see Theorem \ref{thm::convergence} for a formal proof). The number of iterations $q$ is set to 5 in all experiments discussed in this section, which has shown to produce a good compromise between training time and compression ratio.

We notice different trends in the compression ratio for the real and synthetic datasets, which are due to the shorter length of the synthetic trajectories (i.e. shortest paths between a starting and end point) compared to the real ones. In these cases, ONTRAC quickly approaches to its maximum compression ratio.

\begin{figure*}[ht!]
\centering
\subfloat[Syn-I \label{fig::temporal_comp_ratio_syn_one}]{
\includegraphics[width=0.2\textwidth]{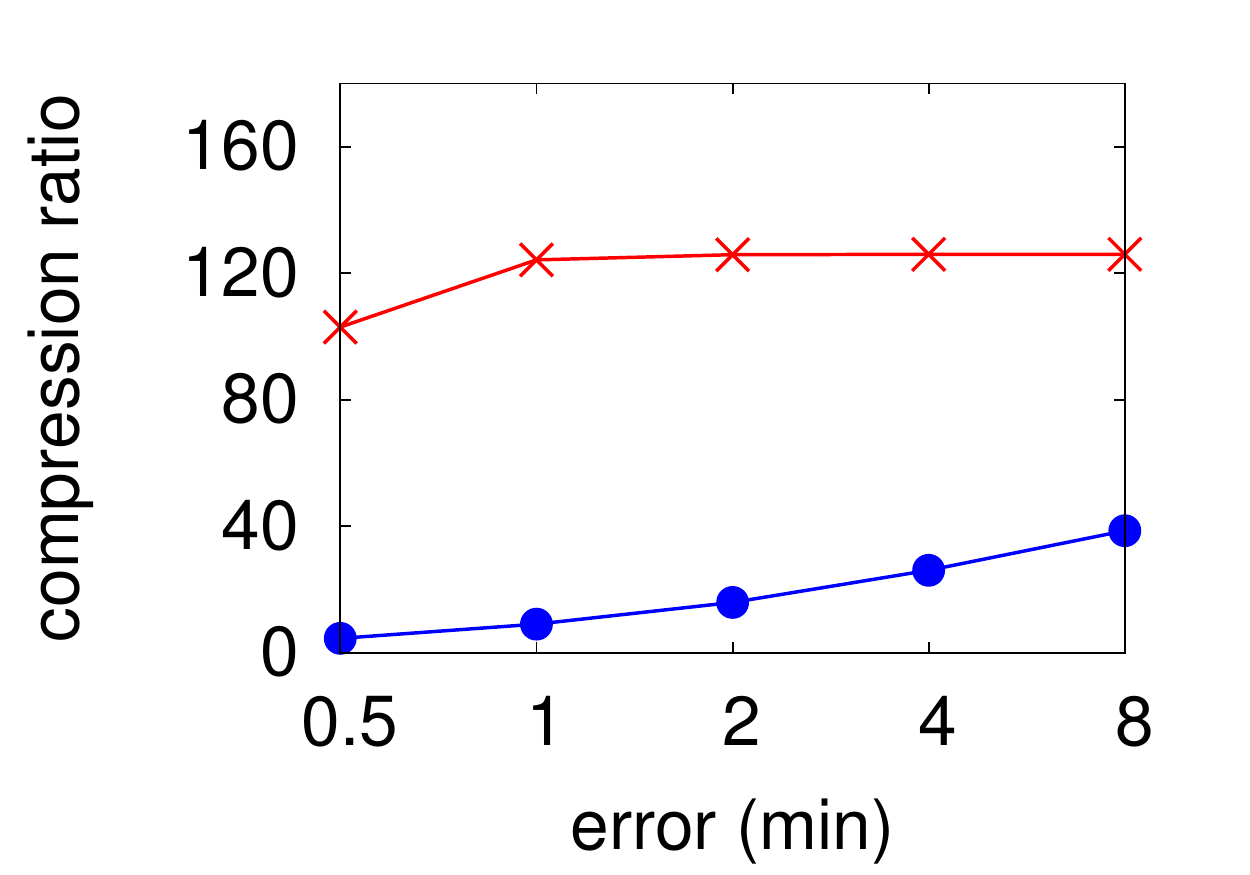}
}
\subfloat[Syn-II \label{fig::temporal_comp_ratio_syn_two}]{
\includegraphics[width=0.2\textwidth]{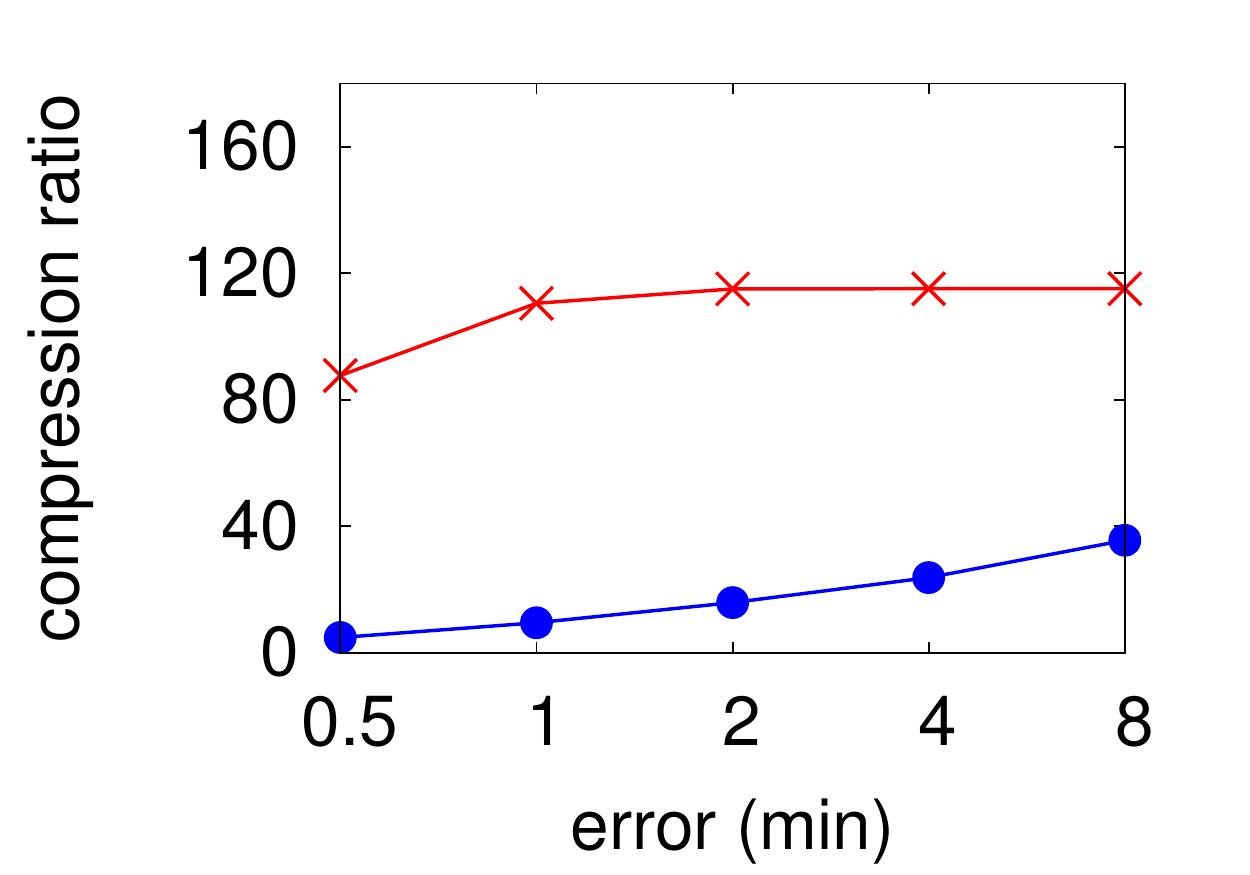}
}
\subfloat[Syn-I \label{fig::temporal_comp_time_syn_one}]{
\includegraphics[width=0.2\textwidth]{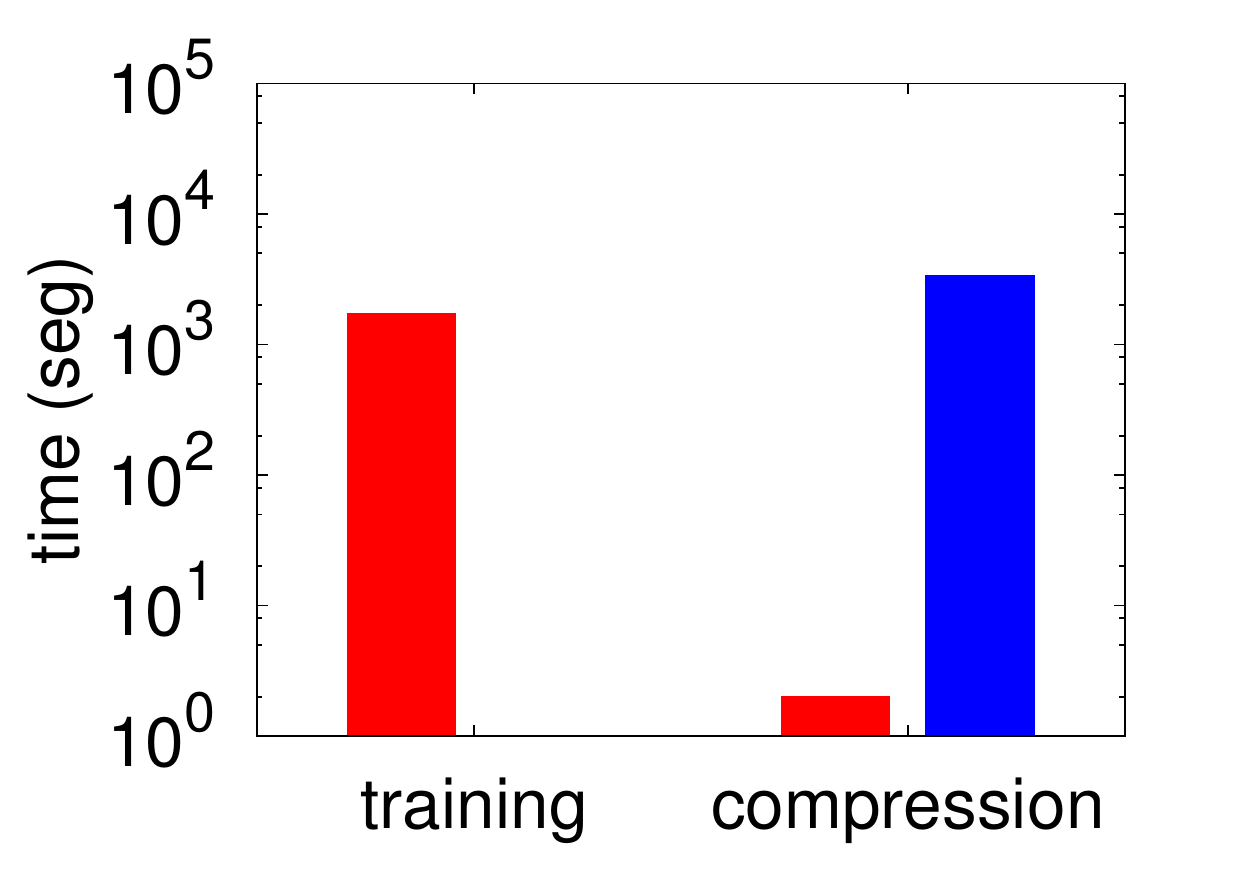}
}
\subfloat[Syn-II \label{fig::temporal_comp_time_syn_two}]{
\includegraphics[width=0.2\textwidth]{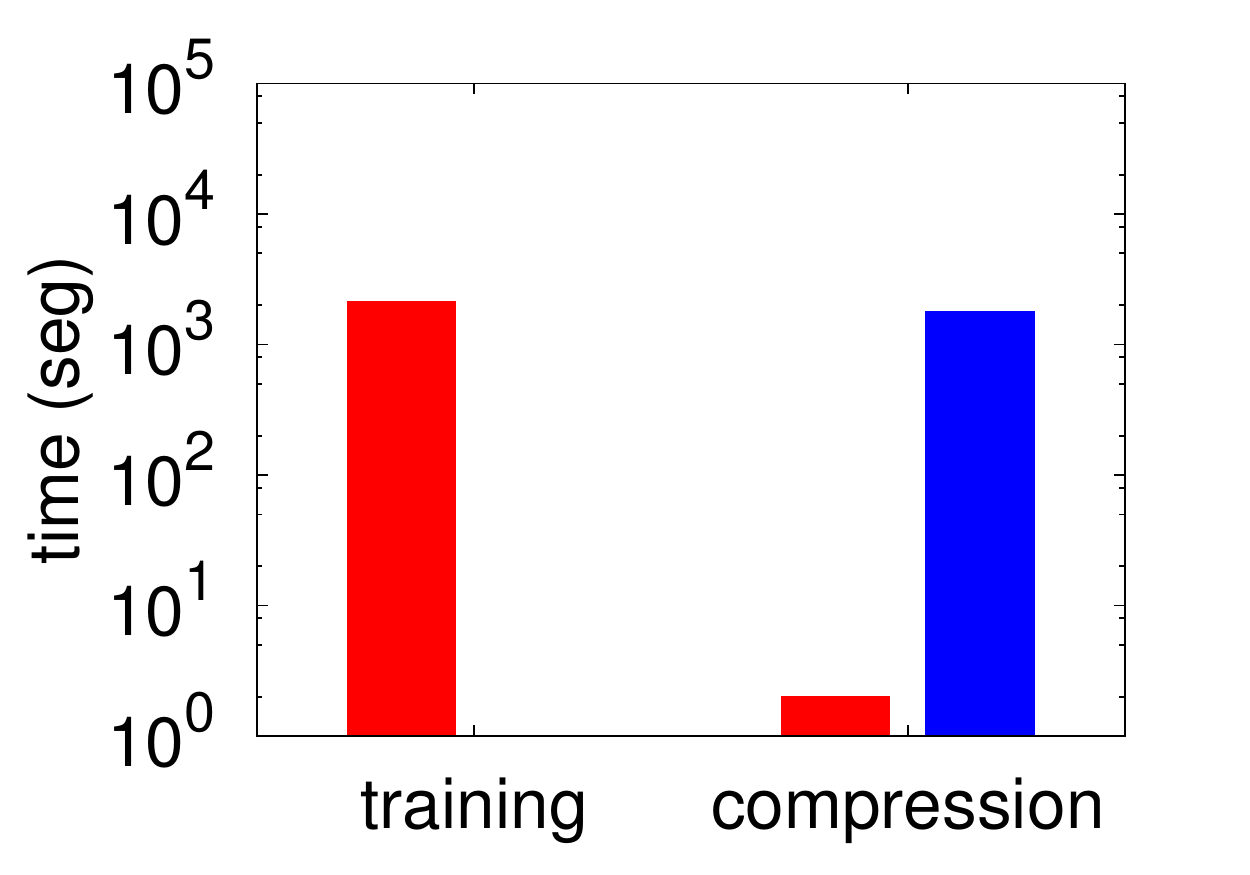}
}
\caption{Temporal compression ratio (a,b), and training and compression times (c,d) for ONTRAC and PRESS using the Syn-I (a,c), and Syn-II (b,d) datasets. }
\end{figure*}

\subsection{Scalability and Querying}

Regarding the synthetic datasets, ONTRAC achieves $50$ and $13$ more inserts/second than the baseline, respectively. These numbers are consistent with the compression ratios presented in Sections \ref{subsec::exp_spatial_compression} and \ref{subsec::exp_temporal_compression}. For synthetic datasets, trajectories are typically short, and thus there is no overhead in decompressing the complete trajectories. 

\begin{figure*}[ht!]
\centering
\subfloat[Syn-I \label{fig::inserts_syn_one}]{
\includegraphics[width=0.2\textwidth]{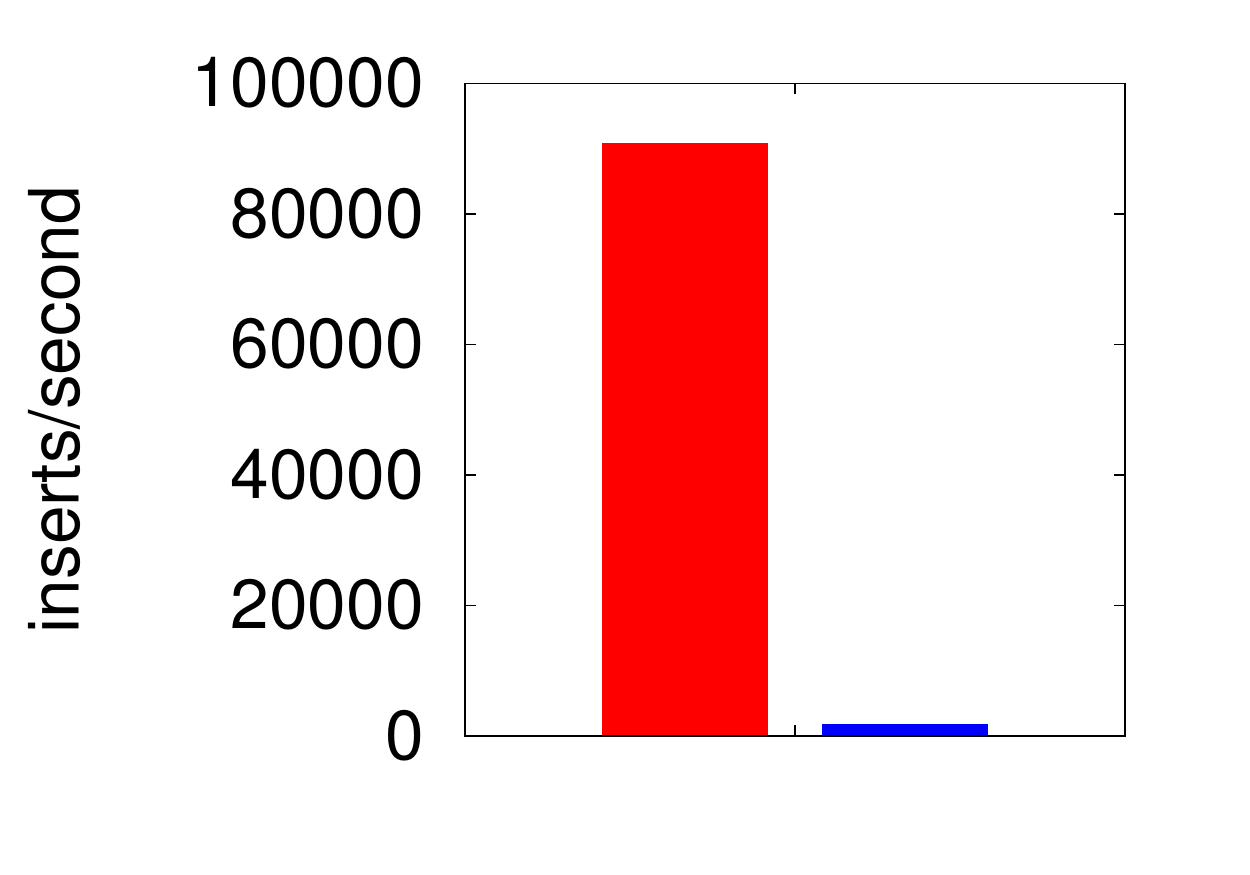}
}
\subfloat[Syn-II \label{fig::inserts_syn_two}]{
\includegraphics[width=0.2\textwidth]{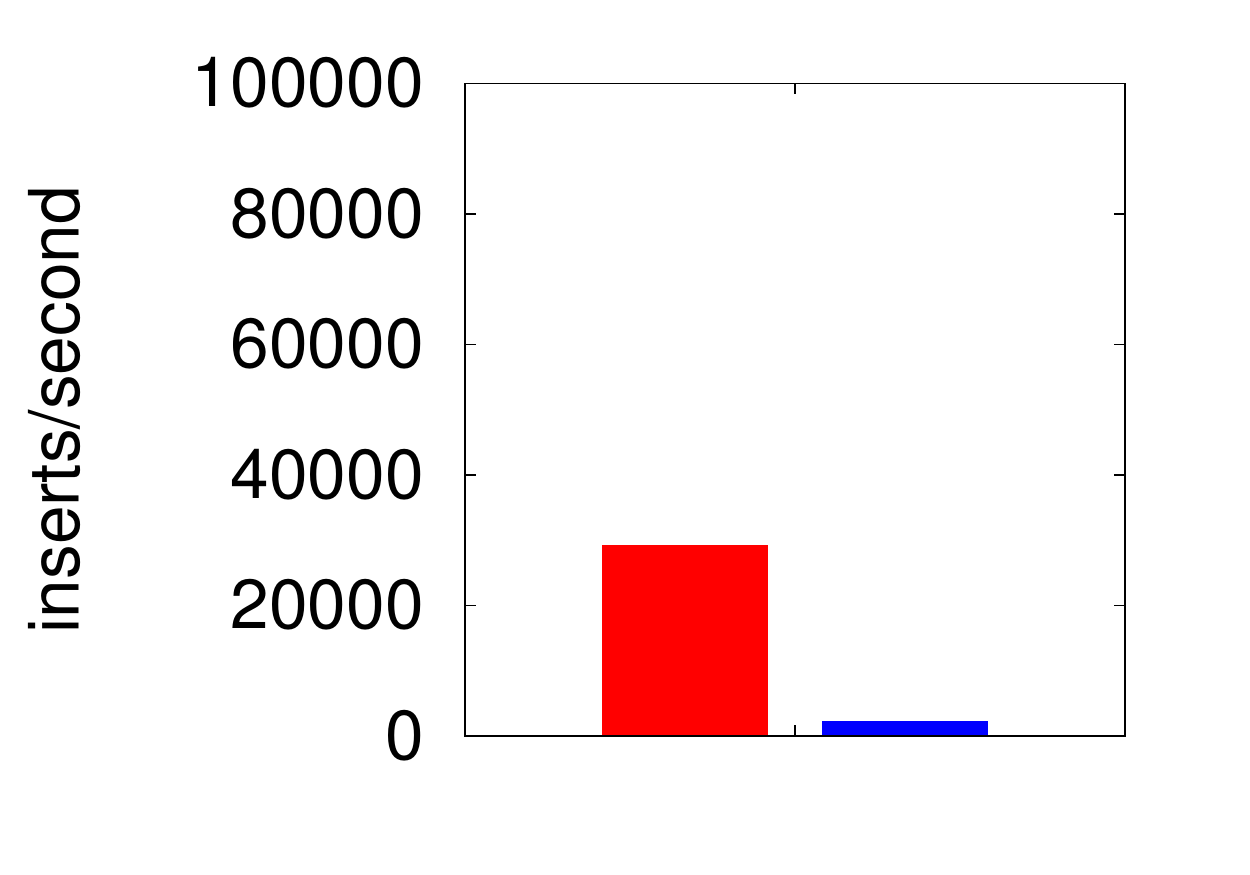}
}
\subfloat[Syn-I \label{fig::query_syn_one}]{
\includegraphics[width=0.2\textwidth]{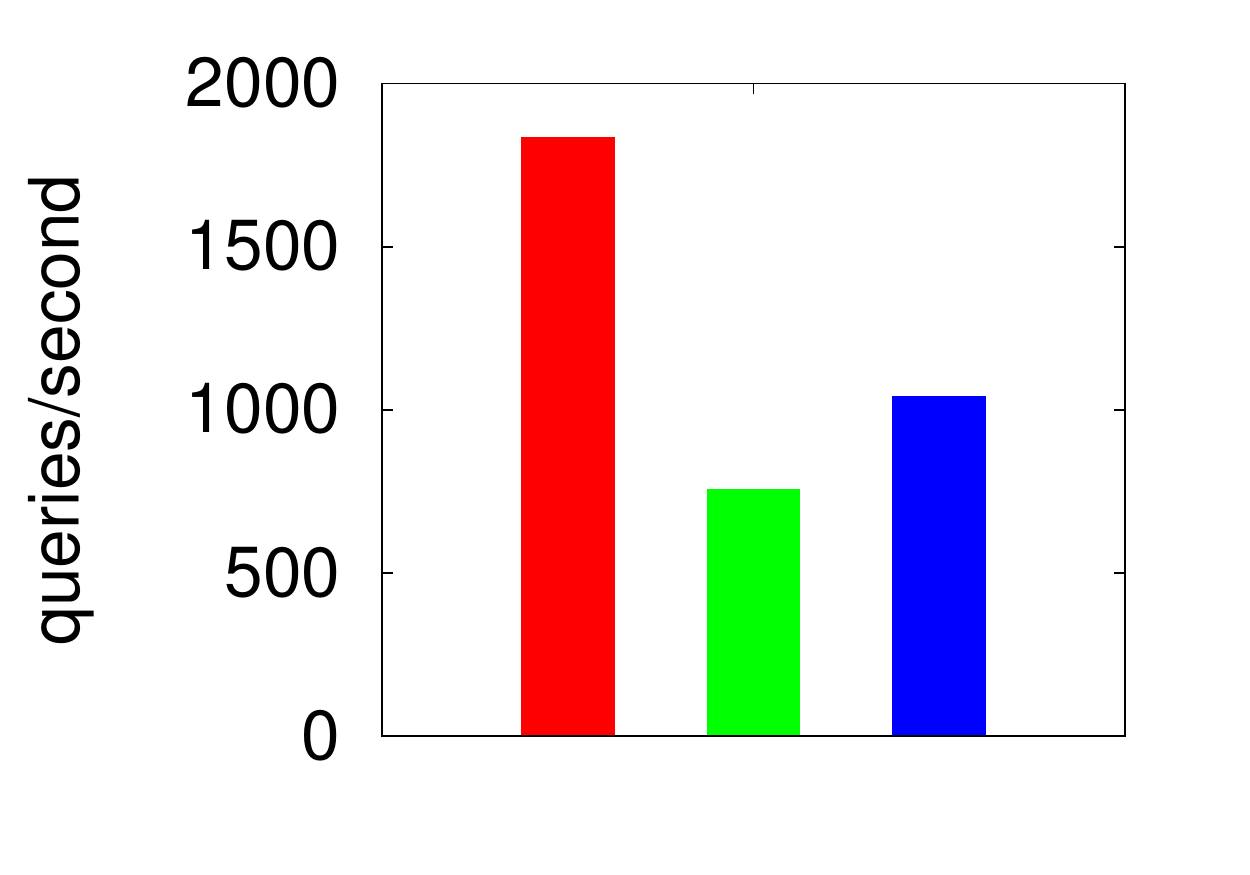}
}
\subfloat[Syn-II \label{fig::query_syn_two}]{
\includegraphics[width=0.2\textwidth]{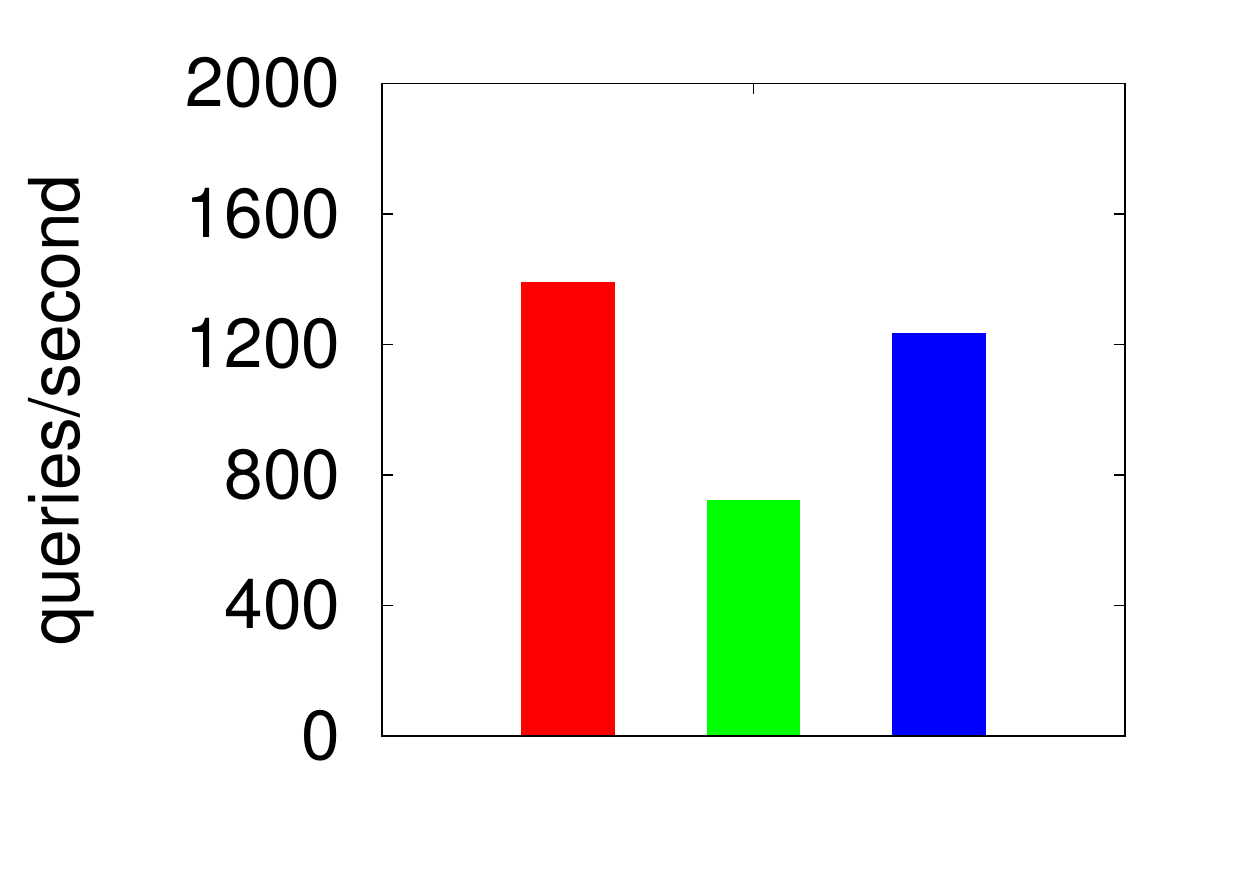}
}
\caption{Inserts (a,b) and queries/second (c,d) when applying ONTRAC with complete (ONTRAC-DC) and partial decompression (ONTRAC-PT) and with no compression (FULL) using Syn-I (a,c) and Syn-II (b,d) datasets.}
\end{figure*}

%% file: traj_comp_long.bbl
\begin{thebibliography}{10}

\bibitem{banerjee2014inferring}
P.~Banerjee, S.~Ranu, and S.~Raghavan.
\newblock Inferring uncertain trajectories from partial observations.
\newblock In {\em ICDM}, 2014.

\bibitem{begleiter2004prediction}
R.~Begleiter, R.~El-Yaniv, and G.~Yona.
\newblock On prediction using variable order markov models.
\newblock {\em JAIR}, 2004.

\bibitem{boyd2004convex}
S.~Boyd and L.~Vandenberghe.
\newblock {\em Convex optimization}.
\newblock 2004.

\bibitem{Cao:2005:NMI:2131560.2131576}
H.~Cao and O.~Wolfson.
\newblock Nonmaterialized motion information in transport networks.
\newblock In {\em ICDT}, 2005.

\bibitem{1096090}
J.~G. Cleary and I.~Witten.
\newblock Data compression using adaptive coding and partial string matching.
\newblock {\em IEEE TCOM}, 1984.

\bibitem{dempster1977maximum}
A.~P. Dempster, N.~M. Laird, and D.~B. Rubin.
\newblock Maximum likelihood from incomplete data via the em algorithm.
\newblock {\em Royal Statistical Society}, 1977.

\bibitem{edward1972likelihood}
A.~Edward.
\newblock {\em Likelihood}.
\newblock J Hopkins University Press, 1972.

\bibitem{feder1994relations}
M.~Feder and N.~Merhav.
\newblock Relations between entropy and error probability.
\newblock {\em IEEE TOIT}, 1994.

\bibitem{gonzalez2008understanding}
M.~C. Gonzalez, C.~A. Hidalgo, and A.-L. Barabasi.
\newblock Understanding individual human mobility patterns.
\newblock {\em Nature}, 2008.

\bibitem{Guo:2014:TTT:2694428.2694432}
C.~Guo, C.~S. Jensen, and B.~Yang.
\newblock Towards total traffic awareness.
\newblock {\em SIGMOD Record}, 2014.

\bibitem{gustafsson_wireless}
F.~Gustafsson and F.~Gunnarsson.
\newblock Mobile positioning using wireless networks: possibilities and
  fundamental limitations based on available wireless network measurements.
\newblock {\em Signal Processing Magazine}, 2005.

\bibitem{hendawi2015predtree}
A.~Hendawi, J.~Bao, M.~Mohamed, and M.~Ali.
\newblock Predictive tree: An efficient index for predictive queries on road
  networks.
\newblock In {\em ICDE}, 2015.

\bibitem{hunter2009path}
T.~Hunter, R.~Herring, P.~Abbeel, and A.~Bayen.
\newblock Path and travel time inference from gps probe vehicle data.
\newblock {\em NIPS}, 2009.

\bibitem{ide2011trajectory}
T.~Id{\'e} and M.~Sugiyama.
\newblock Trajectory regression on road networks.
\newblock In {\em AAAI}, 2011.

\bibitem{kellaris2013map}
G.~Kellaris, N.~Pelekis, and Y.~Theodoridis.
\newblock Map-matched trajectory compression.
\newblock {\em Journal of Systems and Software}, 2013.

\bibitem{kim2007path}
S.-W. Kim, J.-I. Won, J.-D. Kim, M.~Shin, J.~Lee, and H.~Kim.
\newblock Path prediction of moving objects on road networks through analyzing
  past trajectories.
\newblock In {\em KES}, 2007.

\bibitem{liao2003real}
Z.~Liao.
\newblock Real-time taxi dispatching using global positioning systems.
\newblock {\em CACM}, 2003.

\bibitem{neal1998view}
R.~M. Neal and G.~E. Hinton.
\newblock A view of the em algorithm that justifies incremental, sparse, and
  other variants.
\newblock In {\em Learning in graphical models}. 1998.

\bibitem{6799840}
A.~Oran and P.~Jaillet.
\newblock An hmm-based map matching method with cumulative proximity-weight
  formulation.
\newblock In {\em ICCVE}, 2013.

\bibitem{page1999pagerank}
L.~Page, S.~Brin, R.~Motwani, and T.~Winograd.
\newblock The pagerank citation ranking: Bringing order to the web.
\newblock 1999.

\bibitem{piorkowski2009parsimonious}
M.~Pi{\'o}rkowski, N.~Sarafijanovic-Djukic, and M.~Grossglauser.
\newblock A parsimonious model of mobile partitioned networks with clustering.
\newblock In {\em COMSNETS}, 2009.

\bibitem{pvldbSongSZZ14}
R.~Song, W.~Sun, B.~Zheng, and Y.~Zheng.
\newblock Press: A novel framework of trajectory compression in road networks.
\newblock In {\em VLDB}, 2014.

\bibitem{wang2014travel}
Y.~Wang, Y.~Zheng, and Y.~Xue.
\newblock Travel time estimation of a path using sparse trajectories.
\newblock In {\em KDD}, 2014.

\bibitem{xue2013destination}
A.~Y. Xue, R.~Zhang, Y.~Zheng, X.~Xie, J.~Huang, and Z.~Xu.
\newblock Destination prediction by sub-trajectory synthesis and privacy
  protection against such prediction.
\newblock In {\em ICDE}, 2013.

\bibitem{yang2013travel}
B.~Yang, C.~Guo, and C.~S. Jensen.
\newblock Travel cost inference from sparse, spatio temporally correlated time
  series using markov models.
\newblock {\em VLDB}, 2013.

\bibitem{yuan2011driving}
J.~Yuan, Y.~Zheng, X.~Xie, and G.~Sun.
\newblock Driving with knowledge from the physical world.
\newblock In {\em KDD}, 2011.

\end{thebibliography}
